\documentclass[12pt,useAMS, usenatbib]{mn2e}
\usepackage{amsmath}
\usepackage{amssymb}
\usepackage{graphicx}
\usepackage{subfigure}
\usepackage{mathrsfs}

\newcommand{\be}{\begin{equation}} 
\newcommand{\ee}{\end{equation}}
\newcommand{\ba}{\begin{eqnarray}} 
\newcommand{\ea}{\end{eqnarray}}
\newcommand{\D}{{\rm d}}
\newcommand{\Max}{{\rm max}}
\newcommand{\Lim}{{\rm lim}}
\newcommand{\In}{{\rm in}}
\newcommand{\Out}{{\rm out}}
\newcommand{\tot}{{\rm tot}}
\newcommand{\eff}{{\rm eff}}
\newcommand{\oct}{{\rm oct}}
\newcommand{\ain}{a_\In}
\newcommand{\ein}{e_\In}
\newcommand{\aout}{a_\Out}
\newcommand{\eout}{e_\Out}
\newcommand{\emax}{e_\Max}
\newcommand{\emin}{e_{\rm min}}
\newcommand{\elim}{e_\Lim}
\newcommand{\aouteff}{a_{\Out, \eff}}
\newcommand{\vecLin}{{\bf L}_\In}
\newcommand{\vecLout}{{\bf L}_\Out}
\newcommand{\vecLtot}{{\bf L}_\tot}
\newcommand{\hatLin}{\hat{{\bf L}}_\In}
\newcommand{\jm}{j_{\rm min}}
\newcommand{\tk}{t_{\rm k}}
\newcommand{\hatS}{\hat{{\bf S}}}
\newcommand{\hatL}{\hat{{\bf L}}}
\newcommand{\hatJ}{\hat{{\bf J}}}
\newcommand{\msun}{M_\odot}
\newcommand{\rsun}{R_\odot}
\newcommand{\tide}{\rm Tide}
\newcommand{\gr}{\rm GR}
\newcommand{\SL}{{\rm sl}}
\newcommand{\PS}{{\rm ps}}
\newcommand{\A}{\mathscr{A}}

\newcommand{\Mtinunit}{\bar{m}_{01}}
\newcommand{\Mbunit}{\bar{m}_2}
\newcommand{\Mpunit}{\bar{m}_1}
\newcommand{\Msunit}{\bar{m}_0}
\newcommand{\aoutunit}{\bar{a}_{\Out,\eff}}
\newcommand{\ainunit}{\bar{a}_\In}

\newcommand{\Rsunit}{\bar{R}_0}

\def\go{\mathrel{\raise.3ex\hbox{$>$}\mkern-14mu
             \lower0.6ex\hbox{$\sim$}}}

\def\lo{\mathrel{\raise.3ex\hbox{$<$}\mkern-14mu
             \lower0.6ex\hbox{$\sim$}}}

\begin{document}

\title[Eccentricity and Obliquity in Stellar Binaries]{Eccentricity and Spin-Orbit Misalignment in Short-Period Stellar Binaries as a Signpost of Hidden Tertiary Companions} 
\pagerange{\pageref{firstpage}--\pageref{lastpage}} \pubyear{2015}

\label{firstpage}

\author[Kassandra R. Anderson, Dong Lai, \& Natalia I. Storch]{Kassandra R. Anderson$^{1}$\thanks{E-mail: kra46@cornell.edu}, Dong Lai$^{1,2}$, \& Natalia I. Storch$^{3}$ \\ \\ $^{1}$Cornell Center for Astrophysics and Planetary Science, Department of Astronomy, Cornell University, Ithaca, NY 14853, USA \\ $^{2}$Institute for Advanced Study, Princeton, NJ 08540 \\ $^{3}$TAPIR, Walter Burke Institute for Theoretical Physics, Mailcode 350-17, Caltech, Pasadena, CA 91125, USA}
\maketitle

\begin{abstract}
Eclipsing binaries are observed to have a range of eccentricities and spin-orbit misalignments (stellar obliquities).  Whether such properties are primordial, or arise from post-formation dynamical interactions remains uncertain.  This paper considers the scenario in which the binary is the inner component of a hierarchical triple stellar system, and derives the requirements that the tertiary companion must satisfy in order to raise the eccentricity and obliquity of the inner binary.  Through numerical integrations of the secular octupole-order equations of motion of stellar triples, coupled with the spin precession of the oblate primary star due to the torque from the secondary, we obtain a simple, robust condition for producing spin-orbit misalignment in the inner binary:  In order to excite appreciable obliquity, the precession rate of the stellar spin axis must be smaller than the orbital precession rate due to the tertiary companion.  This yields quantitative requirements on the mass and orbit of the tertiary.  We also present new analytic expressions for the maximum eccentricity and range of inclinations allowing eccentricity excitation (“Lidov-Kozai window”), for stellar triples with arbitrary masses and including the non-Keplerian potentials introduced by general relativity, stellar tides and rotational bulges.  The results of this paper can be used to place constraints on unobserved tertiary companions in binaries that exhibit high eccentricity and/or spin-orbit misalignment, and will be helpful in guiding efforts to detect external companions around stellar binaries.  As an application, we consider the eclipsing binary DI Herculis, and identify the requirements that a tertiary companion must satisfy to produce the observed spin-orbit misalignment.
\end{abstract}


\section{Introduction}
Stellar binaries can exhibit a rich variety of dynamical behavior.  In systems with sufficiently small separations, the orbit can precess due to non-Keplerian potentials (e.g. general relativistic corrections), and may also be sculpted by tidal dissipation.  If the binary is a member of a higher multiplicity system, or previously experienced a close encounter with a neighboring star, the orbital properties can be further modified.  In many observed binary systems, whether the orbital elements reflect the properties of the protostellar cloud, or result from post-formation dynamical evolution, remains an open question.  Distinguishing between the two possibilities can shed light into star and binary formation processes.  

A possible signature of post-formation dynamical evolution is stellar spin-orbit misalignment (obliquity).  One method of probing stellar obliquities in binaries is by comparing the inclination of the stellar equator (estimated through measurements of $v \sin i$ and the rotational period) with the orbital inclination.  Using this method, \cite{hale1994} found that solitary binaries tend to have low obliquities when the separation is less than $30-40$ AU, but for separations beyond $30-40$ AU, the obliquities are randomly distributed.  However, for binaries residing in hierarchical multi-systems, even those with small separations can have substantial spin-orbit misalignments, as a result of post-formation dynamical evolution. 

More recently, obliquities have been inferred from measurements of the Rossiter-McLaughlin effect \citep{rossiter1924,mclaughlin1924}.  A handful of eclipsing binaries have orbital axes that are misaligned (in projection) with respect to the spin axis of one or both members.  In the ongoing BANANA Project, an effort to measure obliquities in comparable-mass eclipsing binaries, Albrecht et al. (2007, 2009, 2011, 2013, 2014) present Rossiter-McLaughlin measurements of several systems. Thus far, four systems exhibit spin-orbit alignment \citep{albrecht2007, albrecht2011,albrecht2013}, while two systems contain misaligned components: in DI Herculis both the primary and secondary are misaligned, with $\lambda_{\rm pri} \simeq 72^\circ$ and $\lambda_{\rm sec} \simeq -84^\circ$ \citep{albrecht2009}; in CV Velorum, the primary and secondary have $\lambda_{\rm pri} \simeq -52^\circ$ and $\lambda_{\rm sec} \simeq 3^\circ$ \citep{albrecht2014}. A complementary study of spin-orbit misalignments in unequal mass eclipsing binaries (consisting of FGK-M members) is being undertaken via the EBLM project \citep{triaud2013}.  Although the current sample of binaries with Rossiter-Mclaughlin measurements still consists of only a few members, these efforts, and others (e.g. eclipsing binaries observed by {\it Kepler}, see \citealt{dong2013}), will increase the sample in the coming years.   

In general, it is not clear whether large spin-orbit misalignments in eclipsing binaries are primordial (reflecting the initial state of the protostellar cloud), or have been driven to misalignment due to dynamical interactions with a perturber.  In this paper, we consider the latter scenario, where the eclipsing binary is the inner component of a hierarchical triple stellar system, with a tertiary companion orbiting the center of mass of the inner binary.  If the inclination between the inner and outer orbits is sufficiently high, the eccentricity of the inner binary can undergo periodic excursions to large values, known as Lidov-Kozai (LK) cycles \citep{lidov1962,kozai1962}, see also \cite{harrington1968}.  It is widely believed that binaries with $P_{\rm orb} \lesssim 7$ days are not primordial, but have evolved from wider configurations via LK cycles with tidal friction \citep{mazeh1979, eggleton2001, fabrycky2007,naoz2014}.  Indeed, binaries with periods shorter than this threshold are known to have high tertiary companion fractions [of up to 96 \% for periods $< 3$ days; see \cite{tokovinin2006}], supporting the idea that three-body interactions have played a major role in their formation.  There should also exist a population of longer-period, eccentric binaries that are undergoing LK-driven orbital decay (see \citealt{dong2013}).

It is important to recognize that even a strong perturbation from a tertiary companion on the inner binary does not guarantee the production of spin-orbit misalignment in the inner binary.  If the inner binary achieves a sufficiently small pericenter distance, a torque due to the stellar quadrupole (arising from stellar oblateness) may induce a change in the direction of the spin axis, but the degree of spin-orbit misalignment depends on several factors.  In previous work \citep{storch2014,kra2016}, we have investigated the spin dynamics of a planet-hosting star, as a result of the planet undergoing LK oscillations due to a distant stellar companion \citep[see also][]{storch2015}.  The evolution of the stellar spin-axis can be complicated, with several qualitatively distinct types of possible behavior, depending on the combination of planet mass, stellar spin period and the orbital geometries of the inner and outer binaries.  In particular, for increasingly massive planets ($M_p \gtrsim 5 - 10 M_J$), the coupling between the star and planet can be so strong that spin-orbit misalignment cannot be generated, despite drastic changes in the orbital inclination.  As the mass of the secondary body increases from the planetary to the stellar regime, the ability to generate spin-orbit misalignment is even further hindered.

In light of these previous results, the main goal of this paper is to identify under what circumstances large spin-orbit misalignment can be generated in stellar binaries, due to secular interactions with a tertiary companion.  Tertiary companions can also excite the binary eccentricity.  Another goal of this paper is thus to identify the requirements for a tertiary companion to increase the eccentricity of the inner binary from $e \simeq 0$ to an observed eccentricity $e = e_{\rm obs}$. The results of this paper will help interpret current observations of eclipsing binaries, and guide future efforts to detect tertiary companions in binaries exhibiting large spin-orbit misalignment and/or high eccentricities.

We do not consider the effects of tidal dissipation in this study.  If tidal dissipation is sufficiently strong to circularize the orbit, it will almost certainly align the spin axis with the orbital axis on a shorter timescale, thereby erasing any obliquity excitation due to the outer companion. To avoid this complication, we focus here exclusively on the subset of systems that achieve minimum pericenter distances that are too large for dissipative tides to act.  This is in similar spirit to the focus of the BANANA Project \citep{albrecht2011}. 

This paper is organized as follows.  In Section \ref{sec:lidovkozai}, we review aspects of LK oscillations in hierarchical triples with comparable masses, and including the effects of short-range forces (due to general relativity and tidal and rotational distortion).  This section also contains new results concerning the ``LK window'' of inclinations for eccentricity excitation under general conditions.  In Section \ref{sec:spin_orbit} we discuss the spin-orbit dynamics of binaries undergoing LK cycles, and identify a requirement for generating spin-orbit misalignment.  Section \ref{sec:numerical} presents numerical integrations of the octupole-order secular equations of motion for a large number of triple systems, and compares with the analytic results in Sections \ref{sec:lidovkozai} and \ref{sec:spin_orbit}.  In Section 5, we apply the results to the observed eclipsing binary system DI Herculis, and conclude in Section 6.

\section{Lidov-Kozai Cycles in Triples with Comparable Angular Momentum and Short-Range Forces}
\label{sec:lidovkozai}
\subsection{Setup and Equations}
\label{sec:setup}
We consider a hierarchical triple stellar system, composed of an inner binary with masses $m_0$ and $m_1$, and outer companion with mass $m_2$, orbiting the center of mass of $m_0$ and $m_1$.  In this notation, $m_0$ is the primary body of the inner binary, so that the secondary body always satisfies $m_1 \leq m_0$.  The reduced mass for the inner binary is $\mu_\In = m_0 m_1/m_{01}$, with $m_{01} \equiv m_0 + m_1$.  Similarly, the outer binary has reduced mass $\mu_\Out = m_{01} m_2/m_{012}$ with $m_{012} \equiv m_0 + m_1 + m_2$.   The orbital semi-major axis and eccentricity of the inner and outer binaries are $(a_\In,e_\In)$ and $(a_\Out,e_\Out)$ respectively.  For convenience of notation, we will frequently omit the subscript ``in,'' and define $e = e_{\In}$ and $j = \sqrt{1 - \ein^2}$.  The orbital angular momenta of the inner and outer binaries are denoted by $L_\In$ and $L_\Out$ respectively.

When the inclination between the inner and outer binaries is sufficiently high, the eccentricity and inclination of the inner binary can undergo large, cyclic excursions, known as Lidov-Kozai (LK) oscillations \citep{lidov1962, kozai1962}.  See, for example, Fig.~1 of \cite{holman1997}.  These oscillations are driven by the disturbing potential from the tertiary companion.  To quadrupole order of the potential, the oscillations occur on a characteristic timescale $\tk$ given by
\be
\frac{1}{\tk} = \frac{m_2}{m_{01}} \frac{\ain^3}{\aouteff^3} n,
\label{tk}
\ee
where $n = \sqrt{G m_{01}/\ain^3}$ is the orbital mean motion of the inner binary, and we have introduced an ``effective outer binary separation'' $a_{\Out,\eff}$,
\be
a_{\Out,\eff} \equiv \aout \sqrt{1 - \eout^2}.
\ee

The octupole potential of the outer companion further contributes to the secular dynamics of  the system, introducing under some conditions even higher maximum eccentricities and orbit flipping \citep{ford2000, naoz2013}, as well as chaotic orbital evolution \citep{li2014}.  The ``strength'' of the octupole potential (relative to the quadrupole) is determined by
\be
\varepsilon_{\oct} = \frac{m_0 - m_1}{m_0 + m_1}\frac{\ain}{\aout}\frac{\eout}{1-\eout^2}.
\label{eq:epsilon_oct}
 \ee 
Thus, for equal-mass inner binaries ($m_0 = m_1$), or outer binaries with $\eout = 0$, the octupole contributions vanish.

Additional perturbations on the orbit of the inner binary occur due to short-range-forces (SRFs), including contributions from general relativity (GR), and tidal and rotational distortions of the inner bodies.  These non-Keplerian potentials introduce additional pericenter precession of the inner orbit that acts to reduce the maximum achievable eccentricity \citep[e.g.][]{wu2003,fabrycky2007}, and can suppress the extreme orbital features introduced by octupole-level terms \citep{liu2015}.    

In Section \ref{sec:lidovkozai}, for simplicity, we treat the secondary body in the inner binary ($m_1$) as a point mass (although $m_1$ can be comparable to $m_0$).  As a result, we do not consider the SRFs from tidal and rotational distortion of $m_1$.\footnote{For example, the potential energy due to tidal distortion of $m_1$ is $W_{\tide,1} \sim k_{2,1} G m_0^2 R_1^5/r^6$, while the energy due to tidal distortion of $m_0$ is $W_{\tide,0} \sim k_{2,0} G m_1^2 R_0^5/r^6$, where $k_{2,0}$ and $k_{2,1}$ are the Love numbers of $m_0$ and $m_1$.  For the low mass main-sequence stars of interest in this paper, with $R \propto m^{0.8}$, we have $W_{\tide,1}/W_{\tide,0} \sim (m_1/m_0)^2 \lesssim 1$.}  In order to attain analytical results, for the rest of this section we consider the gravitational potential of the tertiary companion only to quadrupole order (except in Section \ref{sec:coplanar}, where we briefly discuss coplanar hierarchical triples).  These results are thus exact for equal-mass inner binaries ($m_0 = m_1$), or outer binaries with $\eout = 0$.  In Section 4, we perform numerical integrations with octupole included, and including all SRFs (GR, and tidal and rotational distortion in both $m_0$ and $m_1$).

Here we present key results of LK oscillations with SRFs in systems where the angular momenta of the inner and outer binaries are comparable.  The results of this section review and generalize several previous works.  For example, \cite{fabrycky2007} derived the expression for the maximum eccentricity in LK oscillations ($\emax$) with the effects of GR included, in the limit where the angular momentum ratio satisfies $L_\In/L_\Out \to 0$.  \cite{liu2015} presented results for general SRFs (GR, tides and rotational distortion) and general angular momentum ratios.  For $L_\In/L_\Out \ll 1$, they identified the existence of a ``limiting eccentricity'' (see Section \ref{sec:emaxlim}), but for general $L_\In/L_\Out$, \cite{liu2015} did not fully explore the behavior of $\emax$ and the boundaries of parameter space that allow LK oscillations (the ``LK window,'' see Section \ref{sec:window}).  When SRFs are neglected, the equations for general $L_\In/L_\Out$ are first given by \cite{lidov1976} (and rederived by \citealt{naoz2013}), along with the analytical expression for the LK window.  This is further studied by \cite{martin2016} in the context of circumbinary planets.  

The total orbital angular momentum of the system\footnote{ We have neglected the contribution from the spins of $m_0$ and $m_1$, since for stellar parameters of interest in this paper, the spin angular momentum $S$ of each star satisfies $S/L_{\In} \ll 1$.} $\vecLtot = \vecLin + \vecLout$ is constant, with magnitude
\be
L_\tot^2 = L_\In^2 + L_\Out^2 + 2 L_\In L_\Out \cos I, 
\label{eq:ang_mom}
\ee
where $I$ is the mutual inclination between the two orbits.  To quadrupole order, $e_\Out$ and $L_\Out$ are constant.  We can rewrite Eq.~(\ref{eq:ang_mom}) in terms of the conserved quantity $K$, where
\be
K \equiv j \cos I - \frac{\eta}{2} e^2 = \rm {constant},
\label{eq:kozai_constant}
\ee 
and where we have defined
\be
 \eta \equiv \left( \frac{L_\In}{L_\Out}\right)_{e_\In = 0}
= \frac{\mu_\In}{\mu_\Out} \bigg[ \frac{m_{01} a_\In}{m_{012}  a_\Out (1 - e_\Out^2)} \bigg]^{1/2}.
\label{eq:eta}
\ee
In the limit of $L_\In \ll L_\Out$ ($\eta \to 0$), Eq.~(\ref{eq:kozai_constant}) reduces to the usual ``Kozai constant,'' $\sqrt{1 - e^2} \cos I = $ constant.  We will set the initial eccentricity $e_0 \simeq 0$ for the remainder of this paper, so that $K \simeq \cos I_0$.  See Appendix \ref{sec:appendix} for a brief consideration of the initial condition $e_0 \neq 0$.

The total energy per unit mass is conserved, and (to quadrupole order) given by 
\be
\Phi = \Phi_{\rm Quad} + \Phi_{\rm SRF}.
\label{eq:energy_contributions}
\ee
The first term in Eq.~(\ref{eq:energy_contributions}), $\Phi_{\rm Quad}$, is the interaction energy between the inner and outer binaries,
\ba
\Phi_{\rm Quad} & = & -\frac{\Phi_0}{8} \big[ 2 + 3 e^2 - (3 + 12 e^2 - 15 e^2 \cos^2 \omega) \sin^2 I \big] \nonumber \\
& = & -\frac{\Phi_0}{8} \bigg \{ 2 + 3 e^2 - (3 + 12 e^2 - 15 e^2 \cos^2 \omega) \nonumber \\
& & \times \bigg[1 - \frac{1}{j^2}\bigg(K + \frac{\eta}{2}e^2 \bigg)^2 \bigg] \bigg \}.
\label{eq:quad_energy}
\ea
where $\omega$ is the argument of pericenter of the inner binary, and
\be
 \Phi_0 =\frac{G m_2 a_\In^2}{a_{\Out,\eff}^3}.
\ee

The second term in Eq.~(\ref{eq:energy_contributions}), $\Phi_{\rm SRF}$, is an energy term due to short-range forces (SRFs) that lead to additional pericenter precession.  The contributions to $\Phi_{\rm SRF}$  consist of the general relativistic correction, as well as tidal and rotational distortion of $m_0$, so that $\Phi_{\rm SRF} = \Phi_{\rm GR} + \Phi_{\rm Tide} + \Phi_{\rm Rot}$, with \citep[e.g.][]{liu2015}
\ba
\Phi_{\rm GR} & = & - \varepsilon_{\rm GR} \frac{\Phi_0}{j}, \nonumber \\
\Phi_{\rm Tide} & = & - \varepsilon_{\rm Tide} \frac{\Phi_0}{15} \frac{1 + 3 e^2 + (3/8)e^4}{j^9}, \nonumber \\
\Phi_{\rm Rot} & = & - \varepsilon_{\rm Rot} \frac{\Phi_0}{2 j^3},
\label{eq:Phi_SRFs}
\ea
where
\ba
\varepsilon_{\gr} & \simeq & 3 \times 10^{-2}  \frac{\Mtinunit^2 \, \aoutunit^3}{\Mbunit \, \ainunit^4}, \nonumber \\
\varepsilon_{\tide} & \simeq & 9.1 \times 10^{-7} \frac{\bar{k}_{2,0} \, \Mpunit \, \Mtinunit \, \Rsunit^5 \, \aoutunit^3}{\Mbunit \, \Msunit \, \ainunit^8}, \nonumber \nonumber \\
\varepsilon_{\rm Rot} & \simeq & 2.9 \times 10^{-5} \bigg( \frac{P_{*}}{10 \ {\rm d}}\bigg)^{-2} 
\frac{\bar{k}_{q,0} \, \Mtinunit \, \Rsunit^5 \, \aoutunit^3}{\Msunit \, \Mbunit \, \ainunit^5}.
\label{eq:epsilon_SRFs}
\ea
Here, $P_\star$ is the spin period of $m_0$.  The various dimensionless masses and radii, $\bar{m}_i$ and $\bar{R}_i$ are the physical quantities scaled by $\msun$ and $\rsun$.  $\ainunit = \ain/1 \, {\rm AU}$, and $\aoutunit = \aouteff /100 \, {\rm AU}$. $\bar{k}_{2,0}$ is the tidal Love number of $m_0$ scaled by its canonical value $k_{2,0} = 0.03$.  Similarly, $\bar{k}_{q,0}$ depends on the interior structure of $m_0$ and helps quantify the degree of rotational distortion, and is scaled by its canonical value $k_{q,0} = 0.01$ \citep{claret1992}\footnote{$k_{q,0} = (I_3 - I_1)/m_0 R_0^2 \hat{\Omega}_0^2$, where $I_1$ and $I_3$ are the principal moments of inertia, and $\hat{\Omega}_0$ is the spin rate of $m_0$ in units of the breakup rate. $k_{q,0}$ is related to the apsidal motion constant $\kappa$ by $k_{q,0} = 2 \kappa /3$.}. Corresponding terms for the tidal and rotational distortions of $m_1$ are obtained by switching the indices $0$ and $1$ in Eqs.~(\ref{eq:epsilon_SRFs}) (but are neglected in Section 2).

In the expression for $\Phi_{\rm Rot}$ in Eq.~(\ref{eq:Phi_SRFs}), we have assumed alignment of the spin and orbital axes.  When the spin and orbital axes are not aligned, $\Phi_{\rm Rot}$ depends on the spin-orbit misalignment angle.  In this situation, the problem is no longer integrable, and numerical integrations are required (however, see \citealt{correia2015} for an analytic treatment).  In order to attain analytic results, we will assume that the spin and orbital axes are aligned for the remainder of Section 2, and consider the spin-orbit dynamics separately, in Section 4 via numerical integrations. 

For the system parameters of interest in this paper, the GR contribution to the SRFs usually dominates over the rotational contribution at low to moderate eccentricities, and the tidal contribution dominates at very high eccentricities ($e \gtrsim 0.9$).  As a result, $\Phi_{\rm Rot}$ can often be neglected.  This approximation requires that $S \ll L_\In$ (where $S$ is the spin angular momentum of $m_0$), and is always satisfied for the systems considered in this paper.  We also require $\varepsilon_{\rm Rot}/2 j^3 \lesssim 1$ (so that the rotational contribution does not suppress the LK cycles), and $\varepsilon_{\rm Rot}/2 j^3 \lesssim \varepsilon_{\gr}/j$ (so that $\Phi_{\rm Rot} \lesssim \Phi_{\gr}$, i.e. rotational distortion is neglible compared to GR).  Thus, ignoring the effects of rotational distortion is justified for eccentricities that satisfy
\be
1 - e^2 \gtrsim 5.9 \times 10^{-4} \bigg(\frac{\bar{k}_{q,0} \Mtinunit \Rsunit^5 \, \aoutunit^3}{\Msunit \Mbunit \ainunit^5} \bigg)^{2/3} \bigg(\frac{P_\star}{10 \ {\rm d}} \bigg)^{-4/3},
\label{eq:rot_condition1}
\ee
and
\be
1 - e^2 \gtrsim 4.8 \times 10^{-4} \frac{\bar{k}_{q,0} \Rsunit^5}{\Msunit \Mtinunit \ainunit} \left(\frac{P_\star}{10 \ {\rm d}} \right)^{-2}.
\label{eq:rot_condition2}
\ee
Therefore, $\Phi_{\rm Rot}$ is often negligible, unless the spin period is exceptionally rapid, or if the star has a large radius.

For a given initial condition ($I_0$ and $e_0 \simeq 0$), the conservation of $\Phi$ (Eq.~[\ref{eq:energy_contributions}]) and $K \simeq \cos I_0$ (Eq.~[\ref{eq:kozai_constant}]), yield $e$ as a function of $\omega$.  The maximum eccentricity (where $\D e / \D \omega = 0$), is achieved when $\omega = \pi/2$ and $3 \pi / 2$.

\begin{figure*}
\centering 
\includegraphics[width=\textwidth]{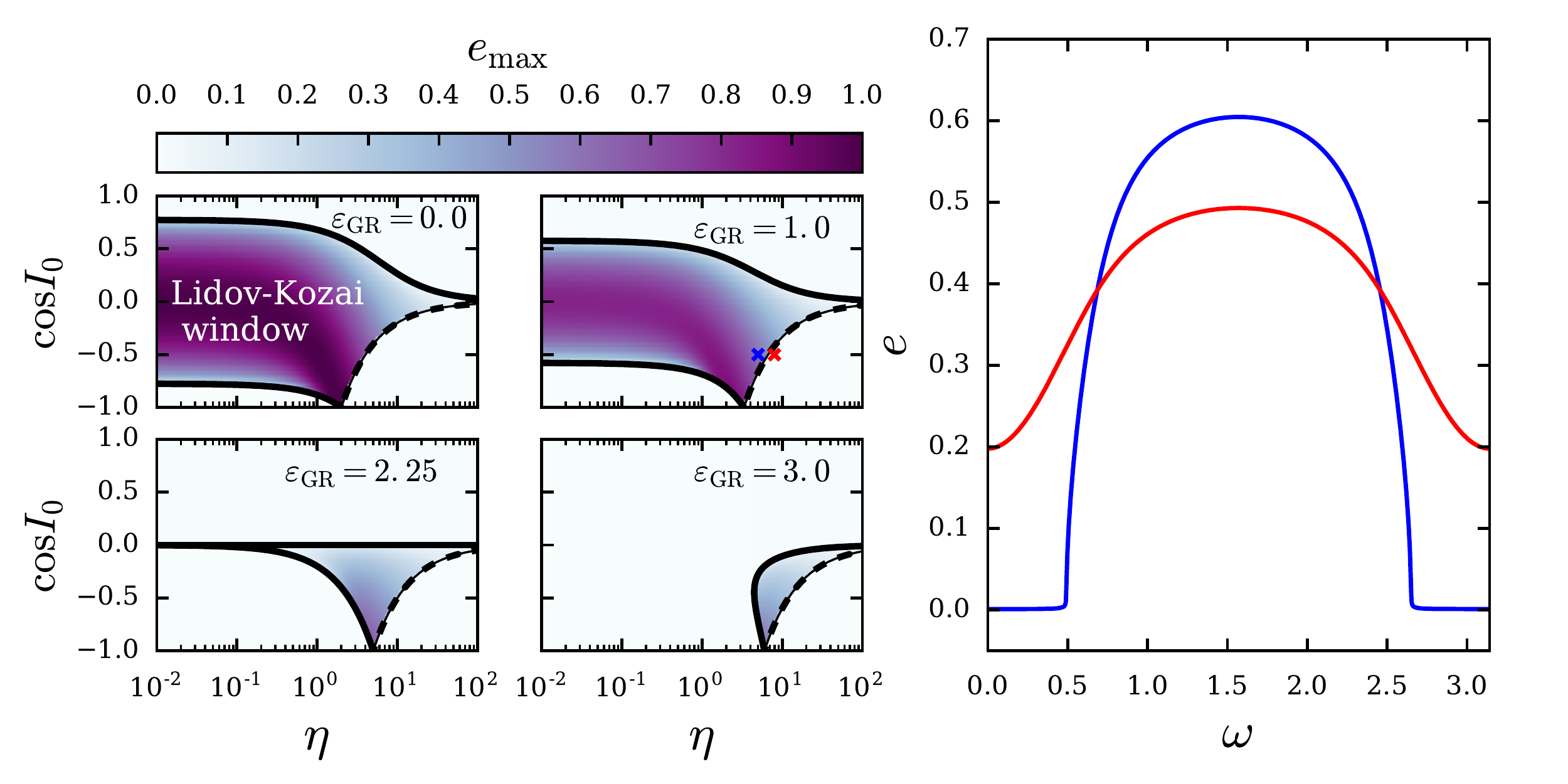}
\caption{{\it Left and center panels}: The ``window'' of inclinations (shaded regions) that allow LK oscillations, versus the angular momentum ratio $\eta$, for various values of $\varepsilon_{\gr}$ (we have set $\varepsilon_{\tide} = \varepsilon_{\rm Rot} = 0$).  The solid lines are obtained from Eq.~(\ref{eq:cos_range}), and the dashed line from Eq.~(\ref{eq:Imin}).  Inside the window, the LK maximum eccentricity is also shown, as calculated in Section \ref{sec:emaxlim}, Eq.~(\ref{eq:energy}).  Combinations of $\cos I_0$ and $\eta$ below the dashed line allow LK eccentricity oscillations, but these oscillations are not connected to the $e_0 \simeq 0$ trajectory. This is illustrated in the rightmost panel, where we show example phase space trajoctories ($\omega, e$) for energies corresponding to the colored crosses in the neighboring uppermost panel (with $\varepsilon_{\gr} = 1.0$).}
\label{fig:window}
\end{figure*}

\subsection{Range of Inclinations Allowing Eccentricity Excitation}
\label{sec:window}
The ``window'' of inclinations allowing LK oscillations (starting from an initial eccentricity $e_0 \simeq 0$) can be determined by enforcing $\emax > 0$.  Expanding for $e^2 \ll 1$, the conservations of energy and $K = \cos I_0$ [valid to $\mathcal{O}(e^6)$] reduce to 
\be
a e^6 + b e^4 + c e^2 = 0,
\label{eq:energy_low_e}
\ee
where
\ba
a & = & \frac{\eta^2}{4} \big(4 - 5 \cos^2 \omega \big) - \frac{\varepsilon_{\gr}}{6} + \frac{5 \varepsilon_{\rm Rot}}{12} + 7 \varepsilon_{\tide} \nonumber \\
b & = & \frac{\eta^2}{4} + (4 - 5 \cos^2 \omega)(1 + \eta \cos I_0) - 1 \nonumber \\ 
& & - \frac{\varepsilon_{\gr}}{3} + \frac{\varepsilon_{\rm Rot}}{2} + \frac{10 \varepsilon_{\tide}}{3} \nonumber \\
c & = &  5 \cos^2 \omega \sin^2 I_0 + 5 \cos^2 I_0 + \eta \cos I_0 - 3 \nonumber \\ 
& & + \frac{4 \varepsilon_{\gr}}{3} + 2 \varepsilon_{\rm Rot} + \frac{4 \varepsilon_{\tide}}{3}. \nonumber \\
& & 
\ea
For $e > 0$, Eq.~(\ref{eq:energy_low_e}) becomes
\be
a e^4 + b e^2 + c = 0.
\label{eq:energy_low_emax}
\ee
This equation determines $e$ as a function of $\omega$ for various parameters $I_0$, $\eta$, $\varepsilon_{\gr}$, $\varepsilon_{\tide}$, and $\varepsilon_{\rm Rot}$.  The maximum eccentricity occurs at $\omega=\pi/2$ and $3\pi/2$.  In order for this $\emax \neq 0$ be reachable from $e_0 \simeq 0$, we require that Eq.~(\ref{eq:energy_low_emax}) admit $e=e_0\simeq 0$ as a solution for some value of $\omega_0\equiv \omega(e_0)$.  Evaluating Eq.~(\ref{eq:energy_low_emax}) at $e = e_0 = 0$ yields
\be
\cos^2 \omega_0 = - \frac{5 \cos^2 I_0 + \eta \cos I_0 - 3 + \varepsilon_{\rm SRF}}{5 \sin^2 I_0},
\ee
where we have defined
\be
\varepsilon_{\rm SRF} \equiv \frac{4}{3} \varepsilon_{\gr} + 2 \varepsilon_{\rm Rot} + \frac{4}{3}\varepsilon_{\tide}.
\ee
Requiring that $\cos^2 \omega_0 \geq 0$ translates into the condition
\be
(\cos I_0)_- \leq \cos I_0 \leq (\cos I_0)_+, 
\label{eq:cos_range}
\ee
where
\be
(\cos I_0 )_{\pm} = \frac{1}{10} \bigg(-\eta \pm \sqrt{\eta^2 + 60 - 20\varepsilon_{\rm SRF} } \bigg).
\label{eq:Icrit}
\ee
In order for $(\cos I_0 )_{\pm}$ to be real, $\eta$ and $\varepsilon_{\rm SRF}$ must satisfy
\be
\eta^2 + 60 - 20 \varepsilon_{\rm SRF} \geq 0.
\label{eq:kozai_condition}
\ee
If $\varepsilon_{\rm SRF} < 3$ then Eq.~(\ref{eq:kozai_condition}) is satisfied for all values of $\eta$.  If $\varepsilon_{\rm SRF} > 3$ and Eq.~(\ref{eq:kozai_condition}) is not satisfied, eccentricity oscillations  cannot be induced for any value of $\cos I_0$. 

Note that while $(\cos I_0 )_{+}$ is less than unity for all values of $\eta$ and $\varepsilon_{\rm SRF}$ (provided that Eq.~[\ref{eq:kozai_condition}] is satisfied), $(\cos I_0 )_{-} > -1$ only when
\be
\eta < 2 + \varepsilon_{\rm SRF} \quad {\rm and} \quad \eta < 10.
\label{eq:cosplus_condition}
\ee
On the other hand, requiring that $\cos^2 \omega_0 \leq 1$ implies that 
\be
\cos I_0 \geq -\frac{2}{\eta} \bigg(1 + \frac{1}{2}\varepsilon_{\rm SRF} \bigg).
\label{eq:Imin}
\ee
Thus, if $\eta > 2 \varepsilon_{\rm SRF}$, then the condition $\cos I_0 \geq (\cos I_0 )_{-}$ (in Eq.~[\ref{eq:cos_range}]) must be replaced by Eq.~(\ref{eq:Imin}).  If $\varepsilon_{\rm SRF} = 0$, the requirement that $\cos I_0 \geq -2/\eta$ is recovered, as identified by \cite{lidov1976}.

The above conditions (Eqs.~[\ref{eq:cos_range}] and [\ref{eq:Imin}]) guarantee that energy conservation Eq.(14) has a physical solution $(e,\omega)=(0,\omega_0)$. Requiring $e^2=\emax^2 > 0$ at $\omega=\pi/2$ implies that $c(\cos\omega=0)<0$, which translates into the condition (\ref{eq:cos_range}).

Figure \ref{fig:window} shows the ``LK window'' of inclinations allowing eccentricity oscillations, determined by Eqs (\ref{eq:Icrit}) and (\ref{eq:Imin}), as a function of $\eta$, for several illustrative values of $\varepsilon_{\gr}$ (and with $\varepsilon_{\tide}, \varepsilon_{\rm Rot}  = 0$).  At moderate eccentricities, the SRF contribution due to GR dominates over the tidal contribution (since $\varepsilon_{\tide} \ll \varepsilon_{\gr}$), and for solar-type stars, GR also dominates over the rotational distortion (since $\varepsilon_{\rm Rot} \ll \varepsilon_{\gr}$).  As a result, adopting the approximation $\varepsilon_{\tide}, \varepsilon_{\rm Rot}  = 0$ is often a valid approximation, except for eccentricities near unity, or for large values of the stellar radius and spin rate, see Eqs.~(\ref{eq:rot_condition1}) and (\ref{eq:rot_condition2}).  

Inside the LK window, the maximum eccentricity is also shown, as calculated in Section \ref{sec:emaxlim}, Eq.~(\ref{eq:energy}). When $\varepsilon_{\gr} = 0$ and $\eta = 0$, the window of inclinations allowing LK oscillations is given by the well known form $- \sqrt{3/5} \leq \cos I_0 \leq \sqrt{3/5}$.  For increasing $\varepsilon_{\gr}$, the window narrows for most values of $\eta$.  When $\varepsilon_{\gr} > 2.25$, the window closes and eccentricity oscillations are completely suppressed for small values of $\eta$.  For larger ($\gtrsim 1$) values of $\eta$, LK oscillations remain possible, but occur only within a very narrow range of inclinations, and are limited to retrograde ($\cos I_0 < 0$) configurations.  We find that for $\varepsilon_{\gr} \gtrsim 5$, the LK window is so narrow for all values of $\eta$, that LK oscillations are for all practical purposes completely suppressed.  The rightmost panel of Fig.~\ref{fig:window} shows phase-space trajectories (contours of constant energy) for two representative points.  The trajectory located just inside the LK window shows that the eccentricity can increase to a large value, starting from $e_0 \simeq 0$. In contrast, the trajectory just outside of the LK window does not connect to $e_0 \simeq 0$.  As a result, for $(\eta,\cos I_0)$ located below the dashed curves in Fig.~\ref{fig:window}, LK oscillations starting from $e_0 \simeq 0$ are completely suppressed.

\subsection{Maximum and Limiting Eccentricities}
\label{sec:emaxlim}
Evaluating the eccentricity at $e_0 = 0$ (where $I = I_0$) and $e = e_{\Max}$ (where $\omega = \pi/2$), allows energy and angular momentum conservation to be expressed as
\be
\begin{split}
& \frac{3}{8} \frac{\jm^2 - 1}{\jm^2} \bigg[ 5 \bigg(\cos I_0 + \frac{\eta}{2} \bigg)^2 - \bigg( 3 + 4 \eta \cos I_0 + \frac{9}{4} \eta^2 \bigg) \jm^2 \bigg. \\ & \bigg. + \eta^2 \jm^4 \bigg] + \bigg( \bigg. \frac{\Phi_{\rm SRF}}{\Phi_0} \bigg) \bigg|_0^{e_\Max} = 0,
\label{eq:energy}
\end{split}
\ee
where $\jm \equiv \sqrt{1 - \emax^2}$.  When the effects of SRFs are negligible, and in the limit $\eta \to 0$, the solution of Eq.~(\ref{eq:energy}) yields the well-known relation $e_\Max = \sqrt{1 - (5/3)\cos^2 I_0}$. Note that the properties of the tertiary companion ($\aout$, $\eout$, $m_2$) enter Eq.~(\ref{eq:energy}) only through the combination $a_{\Out,\eff}/m_2^{1/3}$ and $\eta$. 

\begin{figure}
\centering 
\includegraphics[scale=0.5]{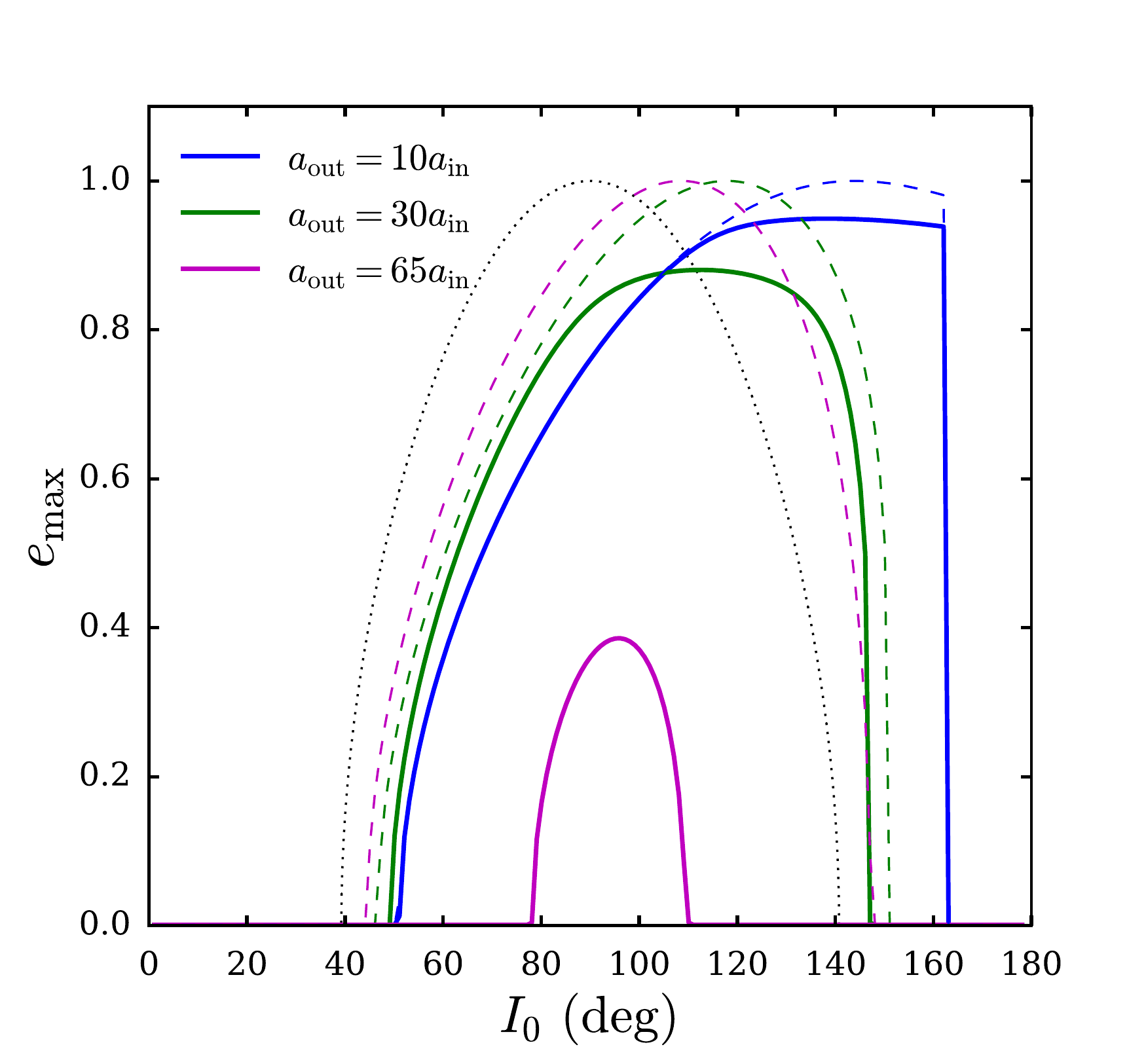}
\caption{The maximum eccentricity of the inner binary, versus the initial inclination $I_0$.  We have fixed $m_0 = m_1 = 1 \msun$, $m_2 = 0.1 \msun$, $a_\In = 0.17$ AU (so that the orbital period is $\sim 15$ days), $\eout = 0$, and varying $a_\Out$, as labeled.  The solid curves show results with SRFs included, and the dashed curves show results without SRFs.  The dotted curve depicts the standard result $e_\Max = \sqrt{1 - (5/3)\cos^2 I_0}$, applicable in the limit $\eta \to 0$ and $\varepsilon_{\gr}, \, \varepsilon_{\rm Rot}, \varepsilon_{\tide} \to 0$.}
\label{fig:emax}
\end{figure}

For general $\eta$, $\varepsilon_{\gr}$, $\varepsilon_{\tide}$, and $\varepsilon_{\rm Rot}$, Eq.~(\ref{eq:energy}) must be solved numerically for $\emax$.    
Fig.~\ref{fig:emax} shows an example of $e_\Max$ versus $I_0$, for an equal-mass inner binary ($m_0 = m_1 = 1 \msun$) with an orbital period of $15$ days, a low-mass outer companion ($m_2 = 0.1 \msun$), and outer binary separations, $a_\Out = 10 \ain, 30 \ain, 65 \ain$ as labeled.

Inspection of Fig.~\ref{fig:emax} reveals that there is a maximum (limiting) achievable value of $\emax$, denoted here as $\elim$, which occurs at a critical initial inclination $I_{0,\Lim}$.  This limiting eccentricity $\elim$ occurs when the initial inclination satisfies the condition ${\rm d} \emax/ {\rm d} I_0 = 0$, or when ${\rm d} j_{\rm min} / {\rm d} I_0 = 0$. Defining $j_{\Lim} \equiv \sqrt{1 - \elim^2}$, and differentiating Eq.~(\ref{eq:energy}) with respect to $I_0$, we find that $I_{0,\Lim}$ is given by
\be
\cos I_{0,\Lim} = \frac{\eta}{2} \left(\frac{4}{5} j_{\Lim}^2 - 1 \right),
\label{eq:Ilim}
\ee    
Obviously, the existence of $I_{0,\Lim}$ requires
$\eta < 2/(1-4 j_{\Lim}^2/5)$.  Notice that $I_{0,\Lim}$ depends on both $\eta$, and on the strength of the SRFs (through $\elim$).  When $\eta \to 0$, $I_{0,\Lim} \to 90^\circ$.  As $\eta$ increases, the critical inclination is shifted to progressively retrograde values ($I_{0,\Lim} > 90^\circ$).

Substituting Eq.~(\ref{eq:Ilim}) into Eq.~(\ref{eq:energy}), we find that the limiting eccentricity $\elim$ is determined by
\be
\frac{3}{8} (j_\Lim^2 - 1) \bigg[ - 3 + \frac{\eta^2}{4} \left(\frac{4}{5} j_\Lim^2 - 1 \right) \bigg] + \bigg. \bigg( \frac{\Phi_{\rm SRF}}{\Phi_0} \bigg) \bigg|_{e = 0}^{e = \elim} = 0.
\label{eq:jlim}
\ee
Equation (\ref{eq:jlim}) may sometimes permit a physical solution for $j_{0,\Lim}$, but imply unphysical values for $\cos I_{0,\Lim}$.  In such cases, $e_\Lim$ cannot be achieved.  As a result, any solution obtained from Eq.~(\ref{eq:jlim}) must also be substituted into Eq.~(\ref{eq:Ilim}) to ensure that $\cos I_{0,\Lim}$ exists.

\begin{figure}
\centering 
\includegraphics[scale=0.65]{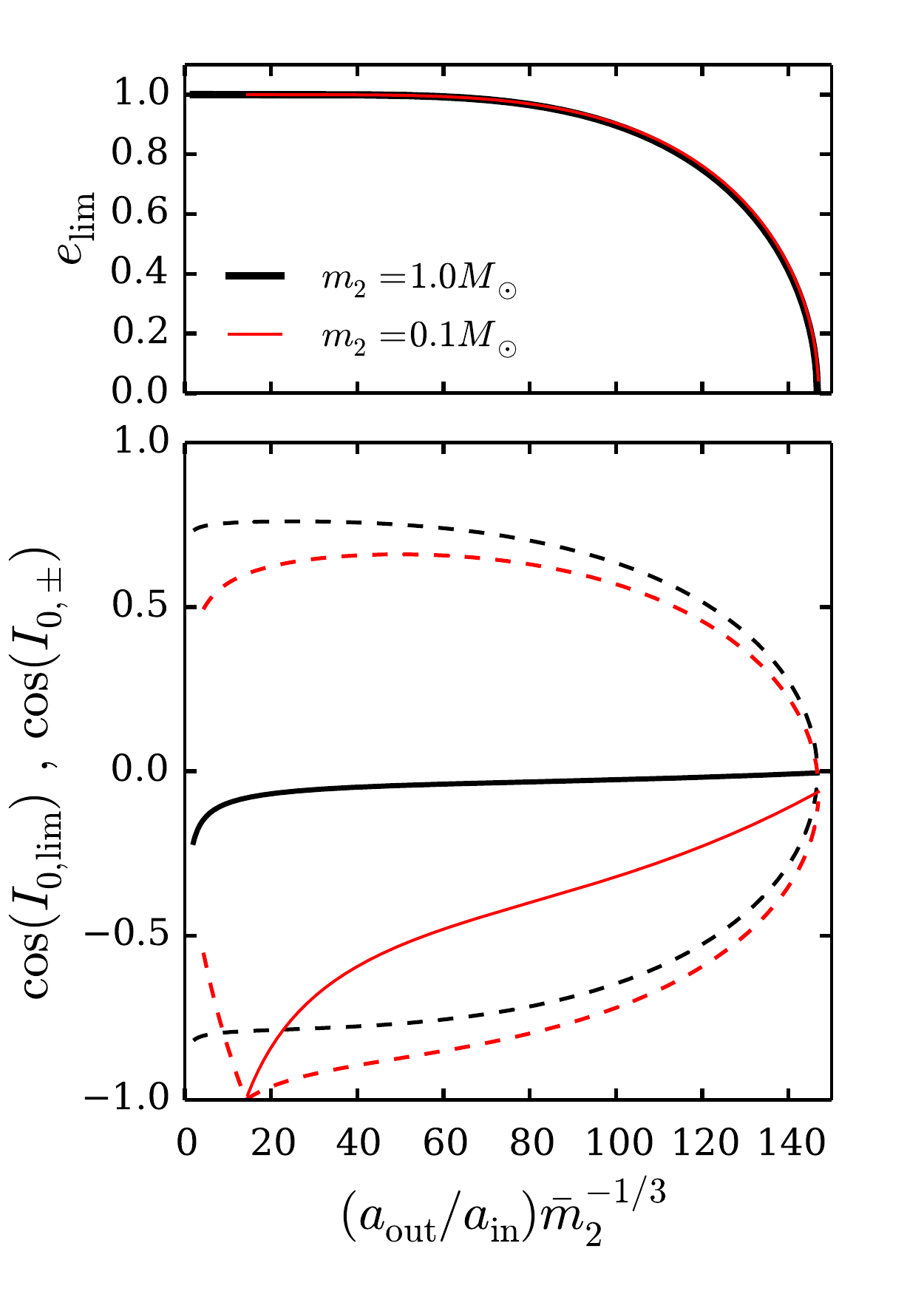}
\caption{Limiting eccentricity $\elim$ and critical inclination $I_{0,\Lim}$, as a function of $(\aout/\ain) \Mbunit^{-1/3}$.  The black curves show $m_2 = 1 \msun$, and the red curves show $m_2 = 0.1 \msun$. The other parameters are $m_0 = m_1 = 1\msun$, $\ain = 0.17$ AU, and $\eout = 0$. In the lower panel, the solid lines indicate $I_{0,\Lim}$, and the dashed lines show the range of inclinations capable of exciting LK oscillations ($I_{{\rm 0},\pm}$), as determined from Eqs.~(\ref{eq:cos_range}) and (\ref{eq:Imin}). As $L_\Out$ decreases relative to $L_\In$ (i.e. $\eta \gtrsim 1$), $I_{0,\Lim}$ is shifted to progressively retrograde values.  For the brown dwarf tertiary, $\cos I_{0,\Lim}$ does not exist for small values of $\aout m_2^{-1/3}$; as a result $e_\Lim$ cannot always be achieved. Notice that $\elim$ is nearly degenerate in terms of $(\aout) \Mbunit^{-1/3}$ (thus the red and black curves nearly coincide in the top panel).}
\label{fig:I0_lim}
\end{figure}

Figure \ref{fig:I0_lim} shows $\elim$ and $I_{0,\Lim}$ as determined from Eq.~(\ref{eq:Ilim}) and Eq.~(\ref{eq:jlim}), along with the ranges of inclinations allowing LK oscillations of any amplitude, from Eqs.~(\ref{eq:cos_range}) and (\ref{eq:Imin}), as a function of $a_{\Out,\eff}/m_2^{1/3}$.  In this example, we have set $\ain = 0.17$ AU and $\eout = 0$, and adopted two values of the tertiary mass: a solar-type perturber ($m_2 = 1 \msun$) and a brown dwarf perturber ($m_2 = 0.1 \msun$). Since Eq.~(\ref{eq:jlim}) depends on $\eta$ only through $\eta^2$, $\elim$ is nearly degenerate in terms of $\aout/m_2^{1/3}$ for the adopted parameters in Fig.~\ref{fig:I0_lim}. For the solar-mass tertiary, $I_{0,\Lim} \simeq 90^\circ$ for all values of $a_{\Out,\eff}$, because $\eta \ll 1$ is always satisfied.  For the brown dwarf tertiary, $I_{0,\Lim} > 90^\circ$ for small values of $a_{\Out,\eff}$, because $\eta \sim 1$.

\subsection{Constraints on Hidden Tertiary Companions from Inner Binary Eccentricities}
\label{sec:hidden}
\begin{figure}
\centering 
\includegraphics[scale=0.55]{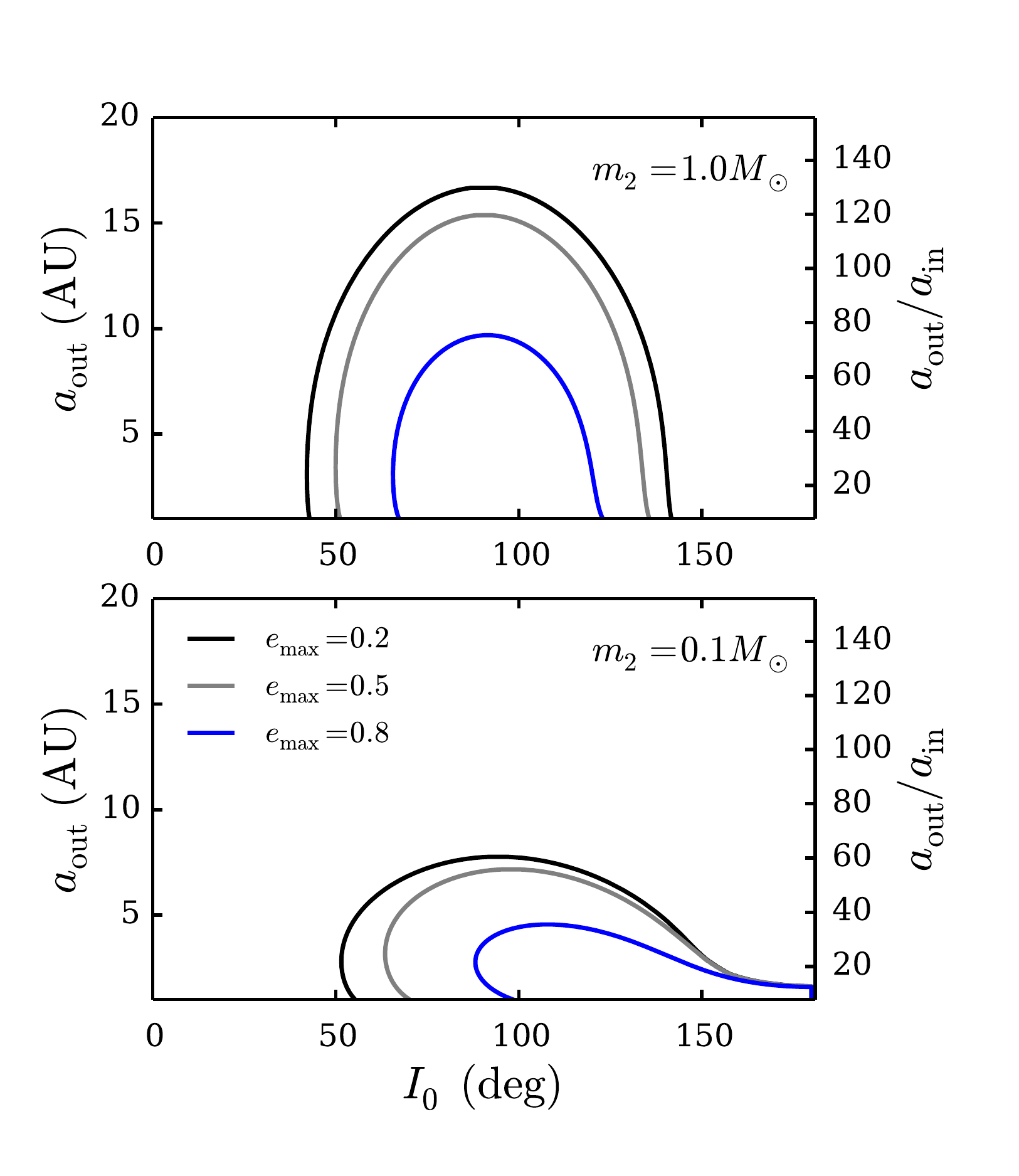}
\caption{Curves in ($I_0, \aout$) parameter space able to produce a given value of $\emax$, as labeled.  For each contour of $\emax$, the region bounded by the curve and the $x$-axis indicates combinations of ($I_0, \aout$) that will yield even higher maximum eccentricities.  Results are shown for a solar-mass outer companion (top), and a brown-dwarf outer companion (bottom).  The inner binary properties are fixed at $m_0 = m_1 = 1 \msun$, $P_{\rm orb} = 15$ days ($\ain = 0.17$ AU), and $\eout = 0$. See also Fig.~\ref{fig:DIHerc} where we show similar calculations applied to the eclipsing binary system DI Herculis.}
\label{fig:aout_vs_I0}
\end{figure}

For an observed binary system with eccentricity $e_{\rm obs}$, we can derive constraints on a possible unseen tertiary companion driving the eccentricity from $e_0 \simeq 0$ to $e = e_{\rm obs}$ through LK cycles. The LK maximum eccentricity must satisfy $\emax \geq e_{\rm obs}$; this places constraints on the mass of the perturber, and the range of mutual inclinations $I_0$ and effective outer separations $a_{\Out,\eff}$.  In Fig.~\ref{fig:aout_vs_I0}, we plot curves of constant $\emax = 0.2, 0.5,0.8$ in $(I_0,\aout)$ space assuming an equal mass inner binary ($m_0 = m_1 = 1 \msun$) with orbital period $P_{\rm orb} = 15$ days, $\eout = 0$, and adopting both solar-type and brown-dwarf perturbers.  The curves were obtained by solving Eq.~(\ref{eq:energy}). For a given $\emax$ contour, the regions inside the curve indicate the parameter space able to produce $e \geq \emax$.  For example, if an observed binary system has $e_{\rm obs} = 0.8$, a solar-mass perturber must be located within $\sim 10$ AU in order to produce the observed eccentricity, and the necessary inclination is restricted to the range $60^\circ \lesssim I_0 \lesssim 120^\circ$.  Similarly, a brown-dwarf companion must be located within $\sim 6$ AU, most likely in a retrograde orbit ($I_0 \gtrsim 90^\circ$).

For $\eta \ll 1$, the properties of the outer perturber required to produce a given eccentricity can be explicity calculated, without having to resort to numerical root-finding in Eq.~(\ref{eq:energy}) or Eq.~(\ref{eq:jlim}).  Neglecting the SRF contribution from rotational and tidal distortion (so that $\varepsilon_{\rm Rot} = \varepsilon_{\tide} = 0$), the LK window (Eq.~[\ref{eq:Icrit}]) is
\be
|\cos I_0 | \leq \frac{1}{5}\sqrt{15 - \frac{20}{3} \varepsilon_{\gr}}.
\ee
Thus, LK oscillations are completely suppressed ($\emax = 0$) when $\varepsilon_{\gr}$ satisfies (see also \citealt{liu2015})
\be
\varepsilon_{\gr} > \frac{9}{4} \left(1 - \frac{5}{3} \cos^2 I_0 \right) \quad {\rm for} \quad \eta \ll 1.
\ee
For an inner binary with specified properties, this translates into a maximum effective perturber distance for LK oscillations (of any amplitude) to occur:
\be
a_{\Out,\eff} < 19.6 \, {\rm AU} \bigg( \frac{\Mbunit}{\Mtinunit^2} \bigg)^{1/3} \bigg( \frac{a_\In}{0.1 \, \rm AU} \bigg)^{4/3} \bigg(1 - \frac{5}{3} \cos^2 I_0\bigg)^{1/3}.
\label{eq:maxaout_zero_ecc}
\ee
Setting $I_0 = I_{0,\Lim} = 90^{\circ}$ yields the absolute maximum effective distance $a_{\Out,\eff}$ for LK oscillations to occur (for any inclination).  

For $\eta \ll 1$, the limiting perturber distance able to drive the eccentricity to $e_{\rm obs}$ can be solved explicitly by setting $\emax = e_{\rm obs} = \elim$, and neglecting the terms in Eq.~(\ref{eq:jlim}) proportional to $\eta^2$,
\be
\begin{split}
a_{\Out,\eff} \simeq & 15.5 \, {\rm AU} \bigg( \frac{a_\In}{0.1 \, {\rm AU}} \bigg)^{4/3} \bigg( \frac{\Mbunit}{\Mtinunit^2} \bigg)^{1/3} \\
& \times \bigg[ \mathcal{F}_1 + \mathcal{F}_2 \frac{\Mpunit \Rsunit^5}{\Msunit \Mtinunit} \left(\frac{a_\In}{0.1 {\rm AU}} \right)^{-4} \bigg]^{-1/3},
\end{split}
\label{eq:maxaout}
\ee
where we have defined
\ba
& & \mathcal{F}_1 = \frac{1}{j_\Lim (j_\Lim + 1)} \\
& & \mathcal{F}_2 = \frac{2.02 \times 10^{-2}}{1 - j_\Lim^2} \left[\frac{1 + 3 e_\Lim^2 + (3/8) e_\Lim^4}{j_\Lim^9} - 1 \right].
\ea
Expanding $\mathcal{F}_1$ and $\mathcal{F}_2$ appropriately, and setting $e_\Lim =  0$, recovers Eq.~(\ref{eq:maxaout_zero_ecc}) evaluated at $I_0 = 90^\circ$.

\begin{figure}
\centering 
\includegraphics[scale=0.5]{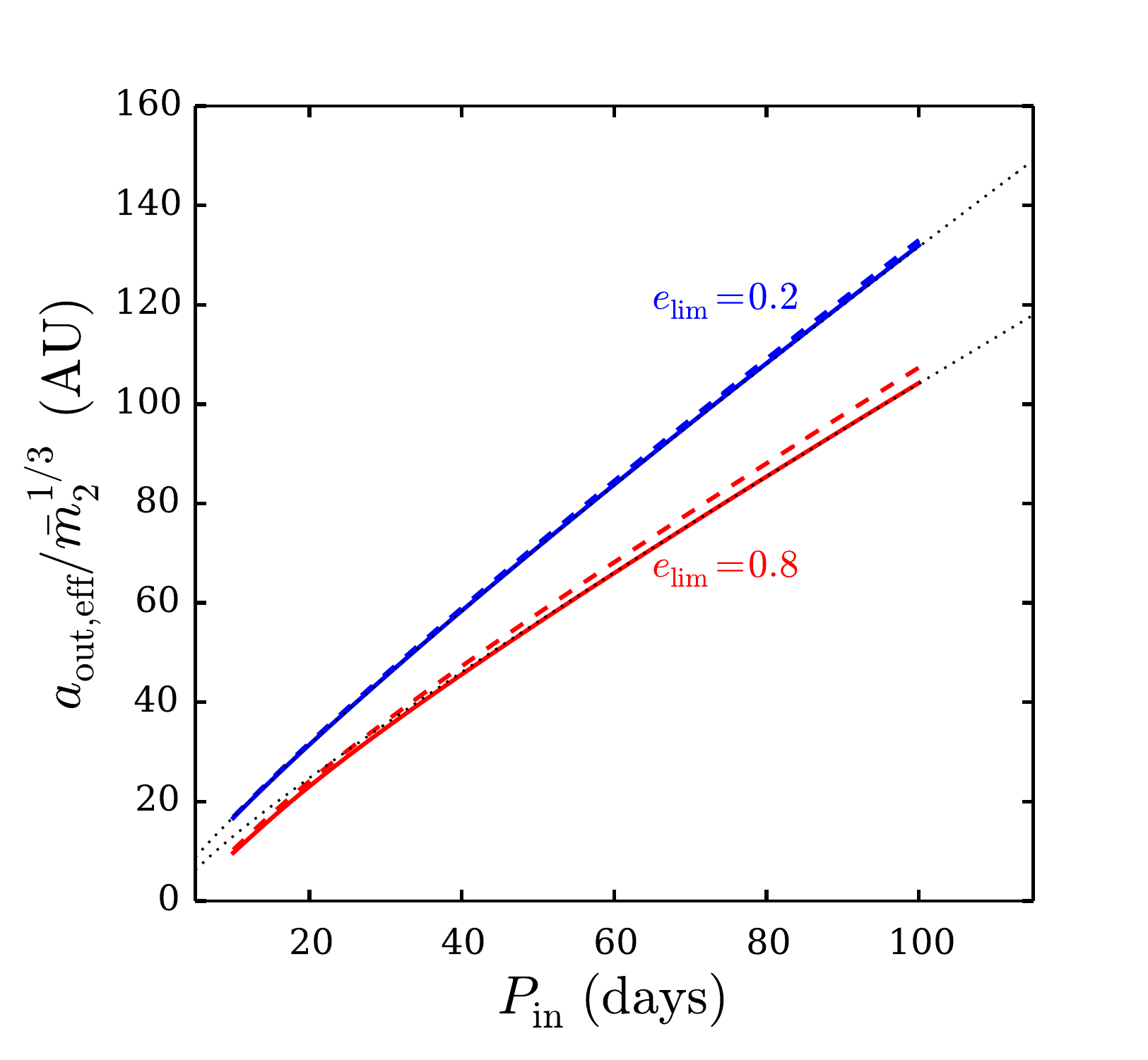}
\caption{Effective perturber distance required to generate a limiting eccentricity $e_{\Lim}$, as labeled, as a function of the inner binary orbital period.  The solid lines depict a solar-mass outer perturber ($m_2 = 1 \msun$), whereas the dashed lines depict a low mass brown dwarf perturber ($m_2 = 0.05 \msun$).  The dashed lines correspond to the expression (\ref{eq:maxaout}), valid in the $\eta \to 0$ limit. For a given inner binary period $P_\In$, in order for an unseen perturber to generate an eccentricity $e_{\rm obs} = 0.2$ ($0.8$), the perturber must have an effective separation lower than the black (blue) value.  Note that the y-axis has been scaled by $(m_2/\msun)^{-1/3}$.}
\label{fig:max_distance}
\end{figure}

In Fig.~\ref{fig:max_distance}, we plot the maximum effective separation required to generate an eccentricity $e_{\rm obs} = 0.2$ and $0.8$, by solving Eq.~(\ref{eq:jlim}).  We also compare this with the approximate ($\eta \ll 1$ limit) expression given in Eq.~(\ref{eq:maxaout}).  The exact solution agrees well with Eq.~(\ref{eq:maxaout}), because the criterion for determining the limiting eccentricity (Eq.~\ref{eq:jlim}) depends on the angular momentum ratio only as $\eta^2$.  Therefore, only when $\eta \to 1$ does the approximate solution deviate from the exact expression.

\subsection{Eccentricity Excitation in Coplanar Systems}
\label{sec:coplanar}
If the inner and outer orbits are coplanar, and the octupole contribution is non-vanishing ($\varepsilon_\oct \neq 0$), the inner and outer binaries can exchange angular momentum, thereby periodically exciting the eccentricity of the inner binary.  In the case of exact coplanarity, the maximum eccentricity can be calculated algebraically \citep{lee2003}.  

The general interaction potential up to octupole order is given in, e.g. \cite{ford2000}, \cite{naoz2013}, and \cite{liu2015}.  If the orbits are exactly coplanar, the interaction energy simplifies to 
\be
\begin{split}
\Phi_{\rm Int} & = \Phi_{\rm Quad} + \Phi_{\rm Oct} \\
& = \frac{\Phi_0}{8} \big[ -2 - 3 e^2 + \frac{15}{8} e (3 e^2 + 4) \varepsilon_\oct \cos \Delta \varpi \big],
\end{split}
\label{eq:coplanar_energy}
\ee
where $\Delta \varpi = \varpi_\In - \varpi_\Out$, with $\varpi$ the longitude of periapsis. 
The total angular momentum $L_{\rm tot} = L_\In + L_\Out$ is also conserved.  
For a given set of orbital geometries (so that both $\Phi$ and $L_{\rm tot}$ are fully specified), $e_\In$ and $e_\Out$ as a function of $\Delta \varpi$ can be obtained.  The maximum value of $e_\In$, $\emax$ occurs at either $\Delta \varpi = 0$ or $\pi$, depending on the initial value of $\Delta \varpi$, and whether $\Delta \varpi$ librates or circulates.  

If either the inner or outer orbit is initially circular, the interaction energy is independent of the initial orientation ($\Delta \varpi$) of the two orbits.  The procedure for calculating $\emax$ is as follows: we specify the initial total energy $\Phi$, including the effects of SRFs $(\Phi = \Phi_{\rm Int} + \Phi_{\rm SRF})$, and the angular momentum $(L_{\rm tot})$, calculate $e$ as a function of $\Delta \varpi$, and determine the maximum value of $e$ \citep[see also][]{petrovich2015b}.  As before, we neglect the contribution to $\Phi_{\rm SRF}$ from rotational distortion ($\Phi_{\rm Rot} = 0$).

\begin{figure}
\centering 
\includegraphics[scale=0.7]{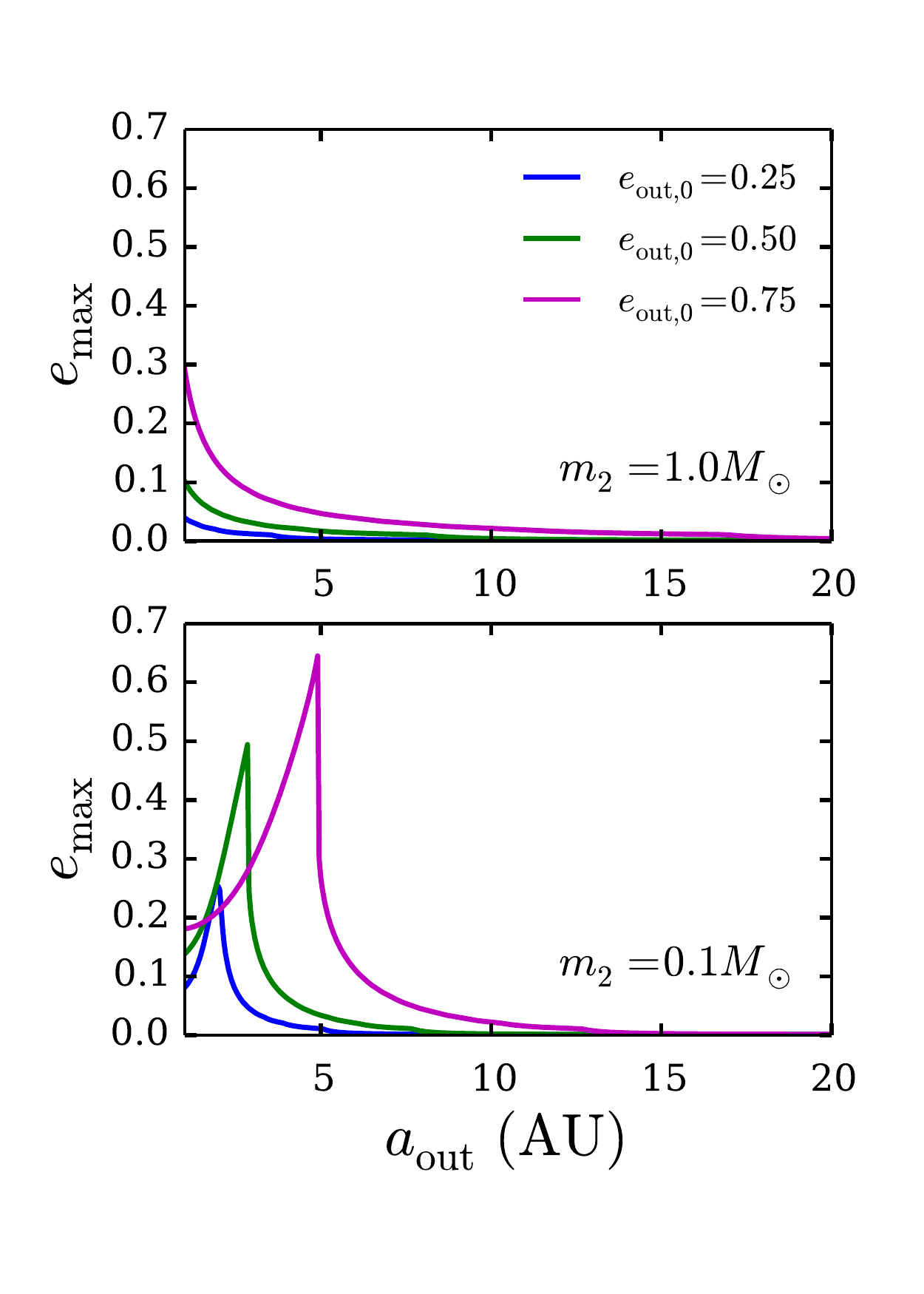}
\caption{Maximum eccentricity $\emax$ for coplanar ($I = 0$) hierarchical triple systems, versus the outer binary semi-major axis.  The properties of the inner binary are fixed, with masses $m_0 = 1 \msun$, $m_1 = 0.5 \msun$, and $P_{\rm orb} = 15$ days.}
\label{fig:coplanar_emax}
\end{figure}

In Fig.~\ref{fig:coplanar_emax} we fix the properties of the inner binary ($m_0 = 1 \msun$, $m_1 = 0.5 \msun$, $P_{\rm orb} = 15$ days), and plot the maximum eccentricity for the two fiducial masses for the perturber ($1 \msun$ and $0.1 \msun$), and varying initial values of $\eout$.  The solar mass perturber must be sufficiently close ($\sim 1 $ AU) and eccentric to excite a substantial eccentricity in the inner binary.  In such configurations, the secular approximation is in danger of breaking down.  The brown dwarf perturber is able to excite higher eccentricities, with a sharp peak.  The sharp peak of $\emax$ at specific value of $\aout$ coincides when the angle $\Delta \varpi$ changes from circulating to librating.  The existence of librating solutions allows for higher maximum eccentricities \citep{lee2003}, and can be understood in terms of an ``apsidal precession resonance'' \citep{liu2015b}.  This ``resonance'' occurs when the apsidal precession of the inner binary (driven by GR and the outer binary) matches that of the outer binary (driven by the inner binary).  However, note that this does not qualify as a ``true resonance'' \citep[see][for further discussion on the nature of this ``resonance'']{laskar1995,correia2010,laskar2012}.

\section{Spin-Orbit Dynamics in Systems Undergoing LK Oscillations}
\label{sec:spin_orbit}
Due to rotational distortion, each member of the inner binary possesses a quadrupole moment, causing a torque and mutual precession of the spin axis ${\bf S}$ and the orbital axis $\vecLin$. Here we discuss the precession of the primary member of the inner binary ($m_0$).  Similar results for the spin precession of $m_1$ are obtained by switching the indices $0$ and $1$ in the following expressions.

The spin axis of $m_0$ precesses around $\hatLin = \hatL$ according to
\be
\frac{\D \hatS}{\D t} = \Omega_{\PS} \hatL \times \hatS,
\label{eq:dSdt}
\ee
where the symbol $\hat{}$ denotes unit vectors, and where the precession frequency $\Omega_\PS$ is given by
\be
\Omega_{\PS} = - \frac{3 G m_1 (I_3 - I_1) \cos \theta_{\SL}}{2 \ain^3 j^3 S}.
\label{OmegaPS}
\ee
In Eq.~(\ref{OmegaPS}), the spin-orbit angle is defined by $\cos \theta_\SL = \hatS \cdot \hatL$, and $I_3 - I_1$ are the principle moments of inertia of $m_0$. \footnote{There is also a spin-spin interaction, of order $G Q_0 Q_1/r^5$, where $Q_{0,1} = (J_2 \, m R^2)_{0,1}$ is the rotation-induced quadrupole moment. This is much smaller than the $S$-$L$ terms, of order $G Q_{0,1} m_{1,0}/r^3$.  In addition, spin-spin resonances may occur when the precession frequencies of the spin axes (Eq.~[\ref{OmegaPS}]) become equal \citep{correia2016}.  However, although this latter effect is captured by our numerical integrations in Section 4, such spin-spin interactions do not play an important dynamical role in the systems of interest here.}

Meanwhile, the orbital axis of the inner binary precesses and nutates around the total orbital angular momentum axis ${\bf J = \vecLin + \vecLout}$, with frequency $\Omega_L = |\D \hatL/\D t|$.  In general, $\Omega_L$ is a complicated function of eccentricity, but takes the approximate form \citep{kra2016}.  
\be
\Omega_L \simeq \frac{3 (1 + 4 e^2)}{8 \tk \sqrt{1 - e^2}} |\sin 2 I |.
\label{OmegaL}
\ee 
Eq.~(\ref{OmegaL}) is exact at $e = 0$ and $e = \emax$. Both $\Omega_\PS$ and $\Omega_L$ are strong functions of eccentricity, and thus can undergo large variation during a single LK cycle.

As described in \cite{storch2014}, the dynamical behavior of $\hatS$ under the influence of a secondary body undergoing LK oscillations depends on the ratio $|\Omega_\PS / \Omega_L|$.  Here we summarize the key aspects of the dynamics (see also \citealt{storch2015,kra2016}):

If $|\Omega_\PS | \ll |\Omega_L|$ throughout the LK cycle, denoted as the ``non-adiabatic regime,'' $\hatS$ cannot ``keep up'' with $\hatL$ as $\hatL$ precesses around $\hatJ$. As a result, $\hatS$ effectively precesses around $\hatJ$, so that $\theta_{\rm sj} \equiv \cos^{-1}( \hatS \cdot \hatJ) \simeq$ constant.  On the other hand, if $|\Omega_\PS | \gg | \Omega_L |$ throughout the LK cycle, denoted as the ``adiabatic regime,'' $\hatS$ ``follows'' $\hatL$, and the spin-orbit angle $\theta_\SL \simeq$ constant.  Finally, if $| \Omega_\PS | \sim | \Omega_L |$ at some point during the LK cycle, the dynamical behavior is complicated due to secular resonances, and chaotic evolution of $\hatS$ can ensue \citep{storch2015}.  We denote this as the ``trans-adiabatic regime.''

In some cases, inclusion of the backreaction torque from the oblate star on the orbit can considerably complicate this simple classification. In particular, our previous work, beginning with \cite{storch2014}, focused on systems in which the secondary member of the inner binary was a planet.  In such cases, $L_\In$ and $S$ are often comparable during the high-eccentricity phases of the LK cycles, and the backreaction torque from the oblate star on the orbit can be significant. In contrast, here we consider a stellar mass secondary body, so that $L_\In \gg S$ is well satisfied.  As a result, the torque on the orbital axis from the oblate star is negligible\footnote{However, note that, although the expression for $d \vecLin / dt$ is negligible here, the oblate star still causes additional pericenter precession of the orbit.}, resulting in simplified behavior.
 
We introduce an ``adiabaticity parameter'' that characterizes the degree to which the stellar spin axis $\hatS$ ``follows'' the precession of $\hatL$ around $\hatJ$, defined as 
\be
\begin{split}
\A & = \bigg |\frac{ \Omega_\PS \, \tk \, j^3}{\cos \theta_\SL} \bigg | \\ & \simeq 58 \frac{\bar{k}_{q,0} \Mpunit \Mtinunit^{1/2} \Rsunit^3}{\bar{k}_\star \Msunit \Mbunit} \bigg( \frac{P_\star}{5 \, {\rm d}} \bigg)^{-1} \bigg( \frac{a_\In}{0.1 \, {\rm AU}} \bigg)^{-9/2} \bigg( \frac{a_{\Out,\eff}}{10 \, {\rm AU}} \bigg)^{3}.
\end{split}
\label{eq:Aparameter}
\ee
In Eq.~(\ref{eq:Aparameter}), $k_\star = S/m_0 R_0^2 \Omega_\star$ describes the mass distribution of $m_0$, which we set to $k_\star = 0.06$ \citep{claret1992}.  See Section \ref{sec:setup} for definitions and canonical values of the other quantities in Eq.~(\ref{eq:Aparameter}).  Since $P_\star$, $a_\In$, $a_{\Out,\eff}$ can all span wide ranges, $\A$ can vary by many orders of magnitude among possible types of hierarchical stellar triples.

Except for the $\sin 2 I$ factor, $\A$ is of order the ratio of $|\Omega_\PS|$ and $|\Omega_L|$, both evaluated at $e = 0$.  Note that the definition (\ref{eq:Aparameter}) differs from the adiabaticity parameter in  \cite{storch2014} and \cite{storch2015}, and in \cite{kra2016}.  This ``fuzziness'' and multiple possible ways in defining such a parameter arises because, from a theoretical point of view, the dynamical behavior of the spin axis relative to $\hatL$ depends on two distinct (but related) parameters, as shown by \cite{storch2016}.  These two parameters relate to the LK-averaged stellar precession rate, and requires a knowledge of $e(t)$ during the LK cycle to evaluate.  For this paper, our goal is to adopt an adiabaticity parameter that is convenient to evaluate for various triple systems, without requiring prior knowledge of $e (t)$. 

If the adiabaticity parameter $\A$ is greater than a critical value $\A_{\rm crit}$, then the system is always in the ``adiabatic regime'' and $\theta_\SL$ will undergo little variation.  As a result, if the inner binary is formed with $\hatS$ and $\hatL$  aligned, then the spin-orbit angle $\theta_\SL$ will remain small for all time.  On the other hand, if $\A \lesssim \A_{\rm crit}$, large spin-orbit misalignment is possible.  In Section \ref{sec:numerical}, we undertake numerical integrations to determine the behavior of the spin-orbit misalignment angle for different values of $\A$, and identify the value of $\A_{\rm crit} \simeq 3$.  

\section{Numerical Experiments}
\label{sec:numerical}
\subsection{Setup and Computational Procedure}

In this section, we present numerical integrations of the full secular equations of motion of hierarchical stellar triples, and examine the maximum achieved eccentricity of the inner binary ($\emax$) and maximum spin-orbit angle ($\theta_{\SL,\Max}$) over the integration timespan.  We include both the quadrupole and octupole terms for the inner and outer orbits, as well as the effects of SRFs on the inner orbit (pericenter precession due to GR, and tidal and rotational distortion of $m_0$ and $m_1$). The full equations of motion can be found in \cite{liu2015}.  In the absence of octupole ($\varepsilon_\oct = 0$), the evolution of the outer orbit consists of precession of the eccentricity vector ${\bf e}_{\Out}$ (with $\eout$ constant), and precession and nutation of $\hat{{\bf L}}_\Out$ around the fixed total angular momentum axis. 

We simultaneously evolve the spin axis $\hatS$ of $m_0$ due to the torque from $m_1$, as well as the spin axis of $m_1$ due to the torque from $m_0$ (Eq.~[\ref{eq:dSdt}]).  We also include the backreaction torques from both spins on the orbit.  Each spin axis is always placed initially parallel to the orbital axis ($\theta_{\SL,0} = 0$).  Both spin periods are given the same initial value ($P_\star$), and held constant throughout the integration.  The spin-behavior of $m_0$ and $m_1$ is qualitatively identical for comparable mass binaries, and we only present results for $m_0$ (but consider the evolution of both spins in the numerical integrations).

Equal mass inner binaries (for which $\varepsilon_\oct = 0$), and unequal mass inner binaries are considered separately, in Sections \ref{subsec:equal} and \ref{subsec:unequal} respectively.  In each case, we adopt a Monte Carlo approach, and generate a large number of systems with the stellar spin periods and orbital parameters uniformly sampled in the following ranges: $P_\star = 1 - 30$ days, $\ain = 0.1 - 1$ AU, $\aout = (10 - 1000) \ain$, $\eout = 0 - 0.9$, and $I_0 = 0^\circ - 180^\circ$.  We conduct separate experiments for a stellar mass perturber ($m_2 = 1\msun$), and a brown dwarf perturber ($m_2 = 0.1 \msun$).  Systems that satisfy any of the following conditions are discarded:  

\begin{enumerate}
\item To ensure stability, systems that do not satisfy 
\be
\frac{\aout}{\ain} > 2.8 \left(1 + \frac{m_2}{m_{01}} \right)^{2/5} \frac{(1 + \eout)^{2/5}}{(1 - \eout)^{6/5}} \left[1 - 0.3 \frac{I_0}{180^{\circ}} \right]
\label{eq:stability}
\ee
are rejected \citep{mardling2001}.  

\item In order to reduce the number of cases where the range of eccentricity variation is low (or where LK oscillations are completely suppressed), systems with limiting eccentricities that satisfy $\elim < 0.3$ are rejected, where $\elim$ is determined by Eq.~(\ref{eq:jlim}).  As discussed in Section \ref{sec:emaxlim}, for specified inner and outer binary properties, $\emax$ depends on the mutual inclination $I_0$, and $\elim$ is the maximum possible value of $\emax$, occurring at a critical inclination $I_{0,\Lim}$.  Due to the full range of inclinations considered ($I_0 = 0^\circ - 180^\circ$), most systems will not be initialized with $I_0 \sim I_{0,\Lim}$, and will satisfy $\emax \ll \elim$.  Requiring that $\elim \geq 0.3$ thus eliminates many systems that will never undergo excursions to high eccentricity.

\item We do not include the effects of tidal dissipation in the inner binary.  This is justifiable because the focus of this paper is on binaries with pericenter distances large enough such that tidal dissipation has not occurred, thereby preserving the initial semi-major axis of the system.  However, some systems do achieve pericenter distances small enough such that changes in both the orbital and spin angular momentum will occur.  As discussed in \cite{kra2016}, the tidal decay rate in a system undergoing LK oscillations (starting from $e_0 \simeq 0$) is reduced by roughly a factor $\sim \sqrt{1 - \emax^2}$ \citep[see also][for a discussion of the orbital decay rate in LK systems]{petrovich2015}. The decay rate of the semi-major axis in a solar-type inner binary undergoing LK oscillations with maximum eccentricity $\emax$ can be approximated by
\be
\begin{split}
\left|\frac{1}{\ain} \frac{\D \ain}{\D t} \right|_{\rm Tide, Lk} & \sim \frac{1.3 \times 10^{-10}}{{\rm yr}} \frac{\Mpunit \Mtinunit \Rsunit^5}{\Msunit \ainunit} \\
& \times \left(\frac{\Delta t_{\rm lag}}{0.1 \ \rm s}\right) \left(\frac{a_F}{0.08 \ {\rm AU}}\right)^{-7},
\end{split}
\ee
(\citealt{kra2016}), where the equilibrium tide model was assumed \citep{darwin1880,singer1968,alexander1973,hut1981}, $\Delta t_{\rm lag}$ is the (constant) tidal lag time, and we have defined
\be
a_F \equiv \ain (1 - \emax^2).
\ee  

The timescale for changing the spin rate of $m_0$ due to tides is roughly
\be
\begin{split}
\bigg| \frac{1}{S} \frac{\D S}{\D t} \bigg|_{\rm Tide,LK}
& \sim \frac{3 \times 10^{-9}}{\rm yr} \frac{\Mpunit^2 \Mtinunit^{1/2} \Rsunit^3}{\Msunit \ainunit^2} \bigg( \frac{P_\star}{10 \, {\rm days}} \bigg)  \\
& \times \bigg( \frac{\Delta t_{\rm lag}}{0.1 \, {\rm s}} \bigg) \bigg(\frac{a_F}{0.08 {\rm AU}} \bigg)^{-11/2}.
\end{split}
\ee  
This also gives the timescale that tidal dissipation changes the spin-orbit misalignment angle. Systems where this timescale is shorter than $\sim 10^9$ years are affected by tides in terms of their stellar obliquities.  We therefore discard systems that achieved $\ain(1 - \emax^2)$ satisfying
\be
\ain (1 - \emax^2) < 0.08 \ {\rm AU}.
\label{eq:min_aF}
\ee
Although this numerical choice is somewhat arbitrary, we have experimented with slightly higher and lower values, and do not find an appreciable effect on our results.  Note Eq.~(\ref{eq:min_aF}) corresponds to rejecting systems that achieve pericenter distances in the range $0.04 \, {\rm AU} \lesssim \ain (1 - \emax) \lesssim 0.08 \, {\rm AU}$. As a result, adopting the rejection condition in Eq.~(\ref{eq:min_aF}) automatically removes systems that are tidally disrupted, i.e. those systems with pericenter distances less than the tidal disruption radius
\be
a (1 - \emax) \lesssim 2.5 R_0 \left( \frac{m_{01}}{m_0} \right)^{1/3} \simeq 0.01 \ {\rm AU}.
\ee
\end{enumerate}

For each combination of $(m_0, m_1)$ and $m_2$, we generate an initial sample of triples large enough such that, after applying the immediate rejection conditions (i) and (ii), $\sim 2000$ systems remain.  We then integrate each system for a timespan $10^3 \tk$ (in Section \ref{subsec:equal}), and $30 \tk /\varepsilon_\oct$ (in Section \ref{subsec:unequal}), and discard any systems that satisfy Eq.~(\ref{eq:min_aF}).  We record the maximum eccentricity ($\emax$), and the maximum spin-orbit angle ($\theta_{\SL,\Max}$) achieved over the entire integration.    

\subsection{Equal Mass Inner Binary}
\label{subsec:equal}
To start, we focus on equal mass inner binaries ($m_0 = m_1 = 1 \msun$), so that $\varepsilon_\oct = 0$.  In this situation, the maximum achievable eccentricity is specified by the algebraic expression Eq.~(\ref{eq:energy}).

After discarding systems that were expected to have undergone tidal dissipation, we are left with 1779 and 1742 systems with a stellar and brown dwarf outer companion respectively.  These systems have initial angular momentum ratios (see Eq.~\ref{eq:eta}) in the range $\eta \sim 0.04 - 0.2$ for the solar-mass tertiary, and $\eta \sim 0.5 - 1.9$ for the brown dwarf tertiary ($m_2 = 0.1 \msun$). Therefore, triples with stellar mass tertiaries can sometimes be qualitatively understood by the test-particle approximation ($\eta = 0$), whereas the brown dwarf tertiary cannot (however, the dynamical effects of the inner orbit on the outer orbit are always included in our numerical integrations, regardless of perturber mass).

As discussed in Section \ref{sec:spin_orbit}, the qualitative behavior of the spin axis of $m_0$, due to the forcing of $m_1$ depends on the ``adiabaticity parameter'' $\A$ (see Eq.~[\ref{eq:Aparameter}]).  When $\A$ is greater than a critical value $\A_{\rm crit}$, the evolution of the spin-axis is strongly coupled to the orbital evolution, and the spin-orbit angle $\theta_\SL \simeq$ constant. Thus, for systems that begin with $\hatS$ and $\hatL$ aligned, generating spin-orbit misalignment requires that $\A \lesssim \A_{\rm crit}$.  Here we identify the numerical value of $\A_{\rm crit}$.  

Results of our numerical integrations are depicted in Fig.~\ref{fig:Acrit_no_oct}.   Given the wide ranges in orbital geometries and stellar spin rates sampled, the maximum eccentricities range from $\emax \simeq 0 - 0.96$, and $\A$ varies by $5-6$ orders of magnitude. The results in Fig.~\ref{fig:Acrit_no_oct} can be qualitatively understood using the arguments presented in Section 3:

\begin{figure*}
\centering 
\includegraphics[width=0.8\textwidth]{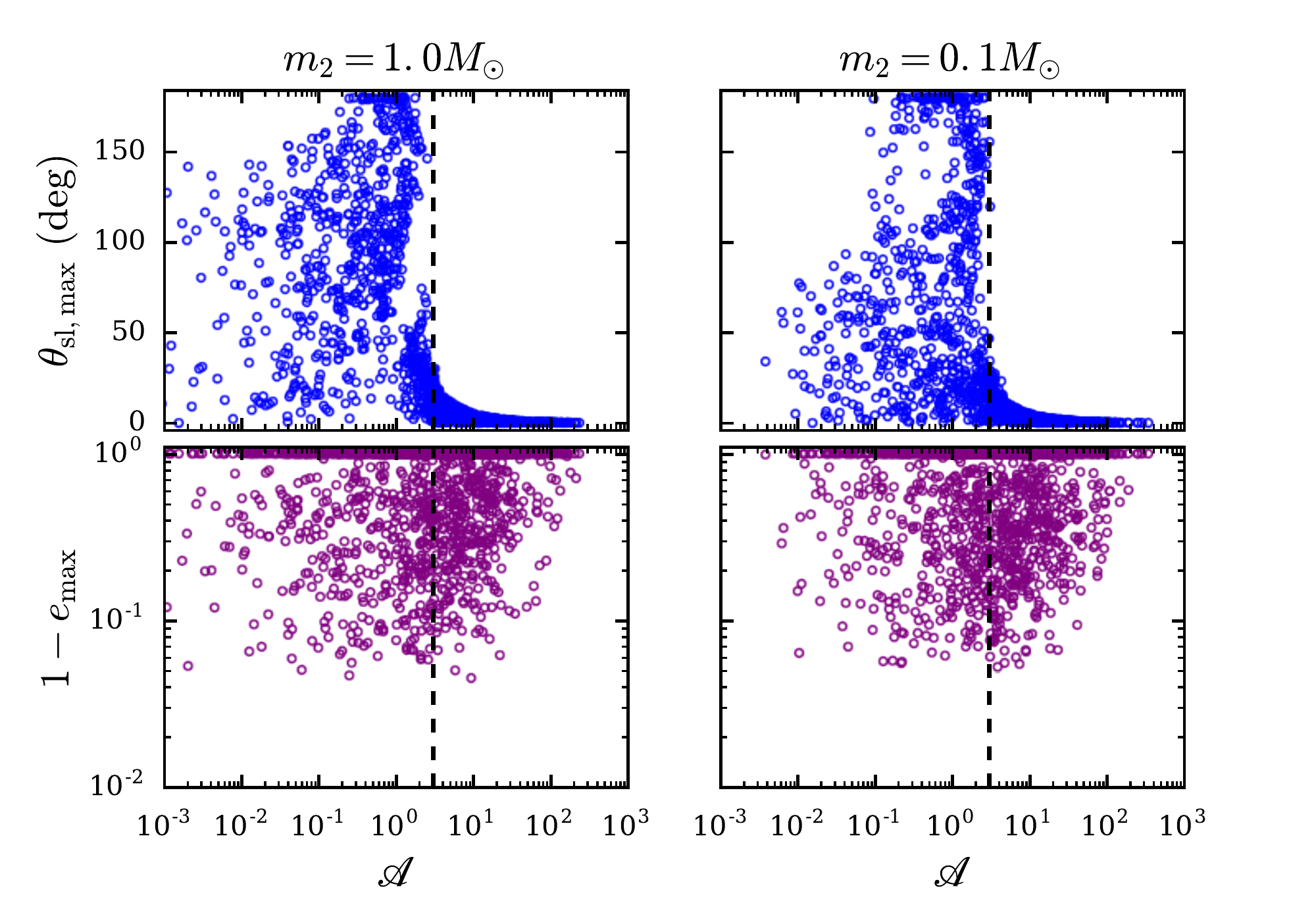}
\caption{Maximum spin-orbit angle $\theta_\SL$ and eccentricity $\emax$ of the inner binary as a function of the adiabaticity parameter, defined in Eq.~(\ref{eq:Aparameter}).  The results are obtained by numerical integrations of systems with an equal mass inner binary ($m_0 = m_1 = 1 \msun$), and other parameters randomly sampled as follows: $P_\star = 1 - 30$ days, $\ain = 0.1 - 1$ AU, $\aout = (10 - 1000) \ain$, $\eout = 0 - 0.9$, and $I_0 = 0^\circ - 180^\circ$.  Lefthand panels show results for a stellar mass ($m_2 = 1 \msun$) tertiary, and righthand panels show results for a brown-dwarf tertiary ($m_2 = 0.1 \msun$). We integrated each system for a period of $10^3 \tk$.  Systems with $\A \gtrsim 3$ maintain low spin-orbit misalignment for the entire integration span (top panels), despite undergoing substantial eccentricity variation (bottom panels).}
\label{fig:Acrit_no_oct}
\end{figure*}

\begin{figure*}
\centering
\includegraphics[width=0.8\textwidth]{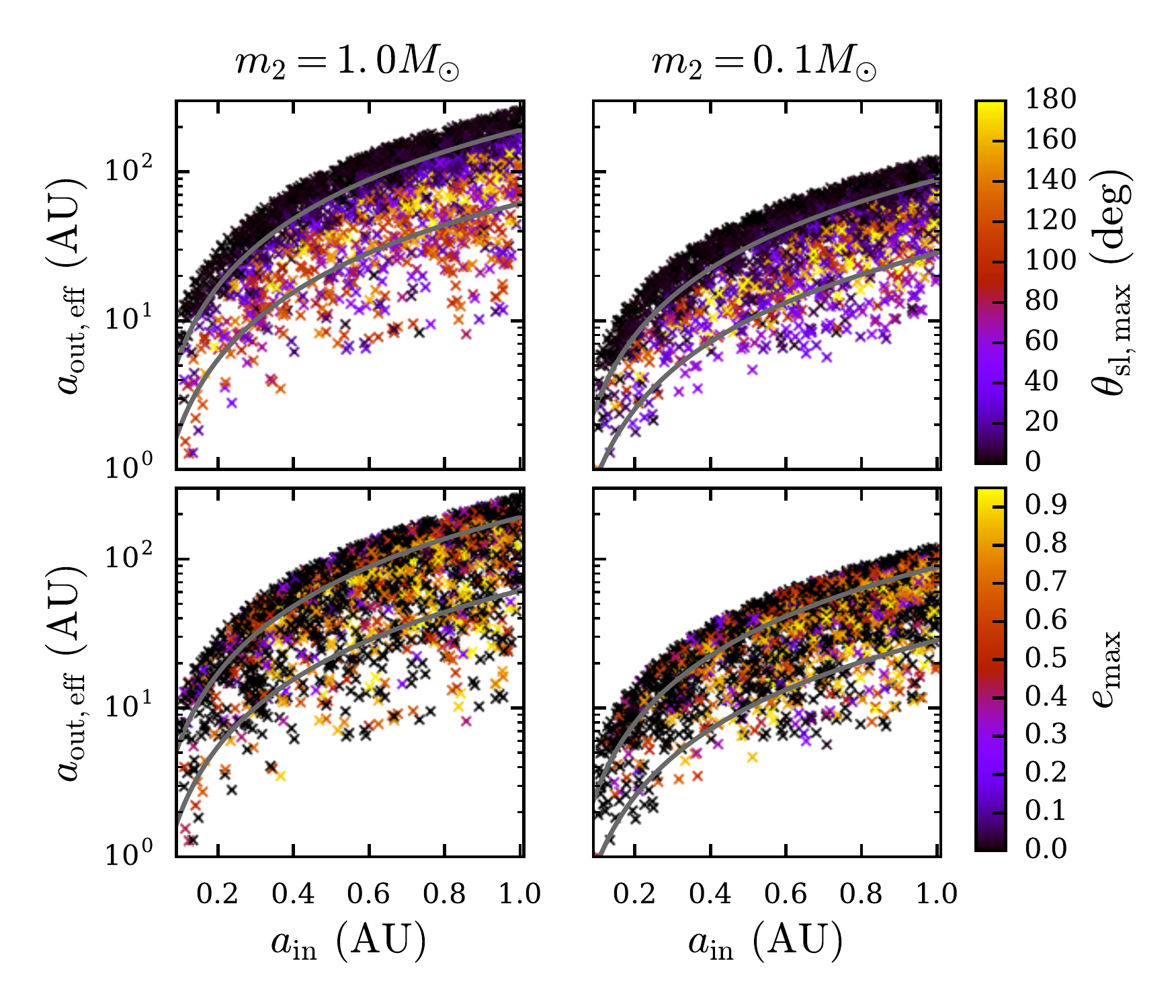}
\caption{Orbital parameters $a_{\Out,\eff} = \aout \sqrt{1 - \eout^2}$ versus $\ain$ for the same sets of triples as in Fig.~\ref{fig:Acrit_no_oct}. The colors indicate the value of $\theta_{\SL,\Max}$ (top panels), and $\emax$ (bottom panels).  We plot curves of constant $\A = 3$ for reference (see Eq.~[\ref{eq:Aparameter}]), with two stellar spin periods selected ($P_\star = 1, 30$ days, grey curves from bottom to top).}
\label{fig:parameters_no_oct}
\end{figure*}

({\it i}) For $\A \lesssim 0.1$, the system is in the non-adiabatic regime (see Section 3), and the precession rate of $\hatS$ around $\hatLin$ is slow compared to the precession of $\hatLin$ around the total angular momentum axis $\hatJ$.  As a result, $\hatS$ effectively precesses around $\hatJ$.  If any nutation of $\hatLin$ relative to $\hatJ$ is neglected, the maximum possible spin-orbit misalignment is approximately $\sim 2 I_0$.  We have confirmed that for $\A \lesssim 0.1$, $\theta_{\SL,\Max} \simeq 2 I_0$.  

({\it ii}) For $0.1 \lesssim \A \lesssim 3$, the evolution of the system is trans-adiabatic (and often chaotic), and $ \theta_{\SL,\Max}$ can momentarily reach $180^\circ$.  

({\it iii}) Systems that satisfy $\A \gtrsim 3$ all maintain low spin-orbit misalignment for the entire integration timespan (with $\theta_{\SL,\Max} \lesssim 30^\circ$).  This is in spite of the fact that many of these systems reached sufficiently high eccentricities (see the bottom panels of Fig.~\ref{fig:Acrit_no_oct}) such that the change in orbital inclination is also large.  Note that the transition from trans-adiabatic to fully adiabatic evolution, in terms of $\A$, occurs abruptly \citep[see also][]{storch2014,storch2015}.  

We conclude from these experiments that a reasonable estimate is $\A_{\rm crit} \simeq 3$.  In order to for substantial spin-orbit misalignment to be generated, the inner and outer binaries must have parameters (i.e. $P_\star, \ain,\aouteff$; see Eq.~[\ref{eq:Aparameter}]) such that $\A \lesssim 3$ is satisfied.

Figure \ref{fig:parameters_no_oct} depicts the results of the same experiments as shown in Fig.~\ref{fig:Acrit_no_oct}, in terms of the parameter space $(a_\In$, $a_{\Out,\eff})$ that we have sampled, with $\theta_{\SL,\Max}$ and $\emax$ indicated by the color.  For a fixed $a_\In$ and $P_\star$, it is clear that in order to generate substantial spin-orbit misalignment, the perturber must have a sufficiently small effective separation $a_{\Out,\eff}$ so that the orbital precession is fast compared to the spin axis precession.  Indeed, from Eq.~(\ref{eq:Aparameter}), the condition $\A \lesssim 3$ translates into
\be
\frac{a_{\Out,\eff}}{\Mbunit^{1/3}} \lesssim 3.7 \, {\rm AU}  \bigg(\frac{\Msunit}{\Mpunit \Mtinunit^{1/2} \Rsunit^3} \bigg)^{1/3} \bigg( \frac{\ain}{0.1 \, {\rm AU}}\bigg)^{3/2}  \bigg( \frac{P_\star}{5 \, {\rm d}}\bigg)^{1/3}.
\ee     
 
\subsection{Unequal Mass Inner Binary: Octupole Results}
\label{subsec:unequal}
Next we consider unequal mass inner binaries, with $m_0 = 1 \msun$ and $m_1 = 0.5 \msun$.  If the octupole potential of the tertiary companion is non-vanishing, i.e. if $\varepsilon_\oct \neq 0$ (which occurs if $m_0 \neq m_1$ and $\eout \neq 0$), the eccentricity of the inner binary can undergo excursions to more extreme values, and under some conditions the orbital inclination can flip (cross $90^\circ$).  The orbital dynamics can be considerably more complicated compared to systems with only the quadrupole potential included.  Here, we examine whether the  results of Section \ref{subsec:equal} remain valid for non-zero $\varepsilon_{\oct}$.

\begin{figure}
\centering 
\includegraphics[scale=0.6]{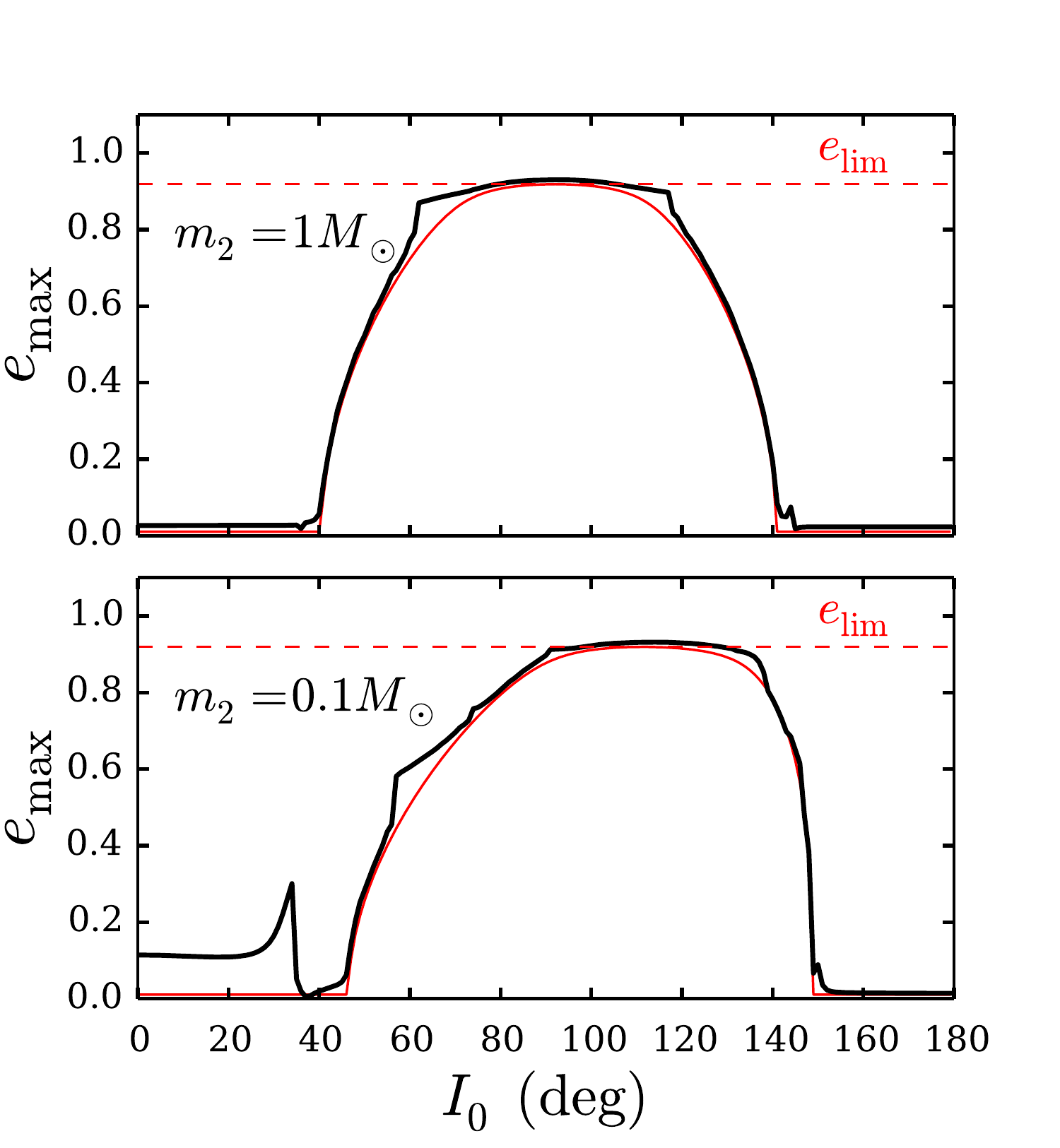}
\caption{Maximum eccentricity $\emax$ achieved over the integration timespan (black curves), compared to the algebraically-determined quadrupole estimate (red curves) from Eq.~(\ref{eq:energy}).  The numerical integrations include quadrupole + octupole contributions, stellar spin-orbit coupling, and all SRFs. Results are depicted for a stellar mass perturber (top panel), and a brown-dwarf perturber (bottom panel).  To illustrate the role of the angular momentum ratio $\eta$ in determining $\emax$, we have fixed $\varepsilon_{\oct} = 0.01$, as well as the quantity $a_{\Out,\eff}/\Mbunit^{1/3} \simeq 6.28$ AU for both panels. The top panels have $\aout \simeq 10.3$ AU, $\eout \simeq 0.79$, and the bottom panels have $\aout \simeq 3.4$ AU, $\eout \simeq 0.51$.  Other parameters (identical for both panels) are: $m_0 = 1 \msun$, $m_1 = 0.5 \msun$, $P_{\rm orb} = 15$ days, $P_* = 10$ days, $\omega_\In = 0$, $\Omega_\In = 0$, $\omega_\Out = 0$.}
\label{fig:elim_example}
\end{figure}

First, we show how the maximum eccentricity is affected. With $\varepsilon_\oct \neq 0$, $\emax$ is no longer specified by Eq.~(\ref{eq:energy}), and determining $\emax$ always requires full numerical integrations.  \cite{liu2015} showed that when considering systems with octupole and SRFs, the maximum achieved eccentricity $e_{\rm max}$ depends on both $I_0$ and $\varepsilon_\oct$, but that $e_{\rm max}$ does not exceed the quadrupole limiting eccentricity $\elim$, as determined by Eq.~(\ref{eq:jlim}).  In other words, even with octupole included, $e_{\rm max} \leq \elim$.  We have confirmed this finding through numerical integrations of the full secular equations of motion (including SRFs).  To demonstrate, Fig.~\ref{fig:elim_example} shows the maximum achieved eccentricity over the integration timespan versus the initial inclination \citep[see also][for similar results]{liu2015}, for the two fiducial values of the perturber mass.  In these examples, the inner binary orbital period is fixed at $P_\In = 15$ days, and the orbital parameters chosen so that $\varepsilon_{\oct} = 0.01$, and $a_{\Out,\eff}/\Mbunit^{1/3} \simeq 6.28$ AU.  We confirm that $\emax$ can have a complicated dependence on $I_0$, especially if $\eta \sim 1$ (bottom panel).  In Fig.~\ref{fig:elim_example}, $\emax$ at $I_0 = 0$ can be calculated using the result of Section 2.4.  The spike in the lower panel (around $I_0 \sim 30^\circ$) may be the result of a secular resonance, but a detailed characterization is beyond the scope of this paper. In general, the degree of deviation of $\emax$ (with octupole) from the quadrupole prediction depends on $\varepsilon_\oct$, as well as on the relative ``strengths'' of the SRFs ($\varepsilon_{\gr}$, and $\varepsilon_{\tide}$) \footnote{Although the effects of SRFs generally suppress $\emax$, under some circumstances, including the effects of GR precession can give rise to eccentricity excitation, yielding $\emax$ that is much higher relative to the case without GR precession included \citep{ford2000,naoz2013b}.}.  We do not attempt to characterize this behavior here (see \citealt{liu2015} for such a characterization in the test-mass limit [$m_1 \ll m_0, m_2$]), and simply present Fig.~\ref{fig:elim_example} as illustrative examples.  Despite the complicated dependence of $\emax$ on inclination, Fig.~\ref{fig:elim_example} shows that $\emax$ does not exceed $\elim$. 

\begin{figure}
\centering 
\includegraphics[scale=0.65]{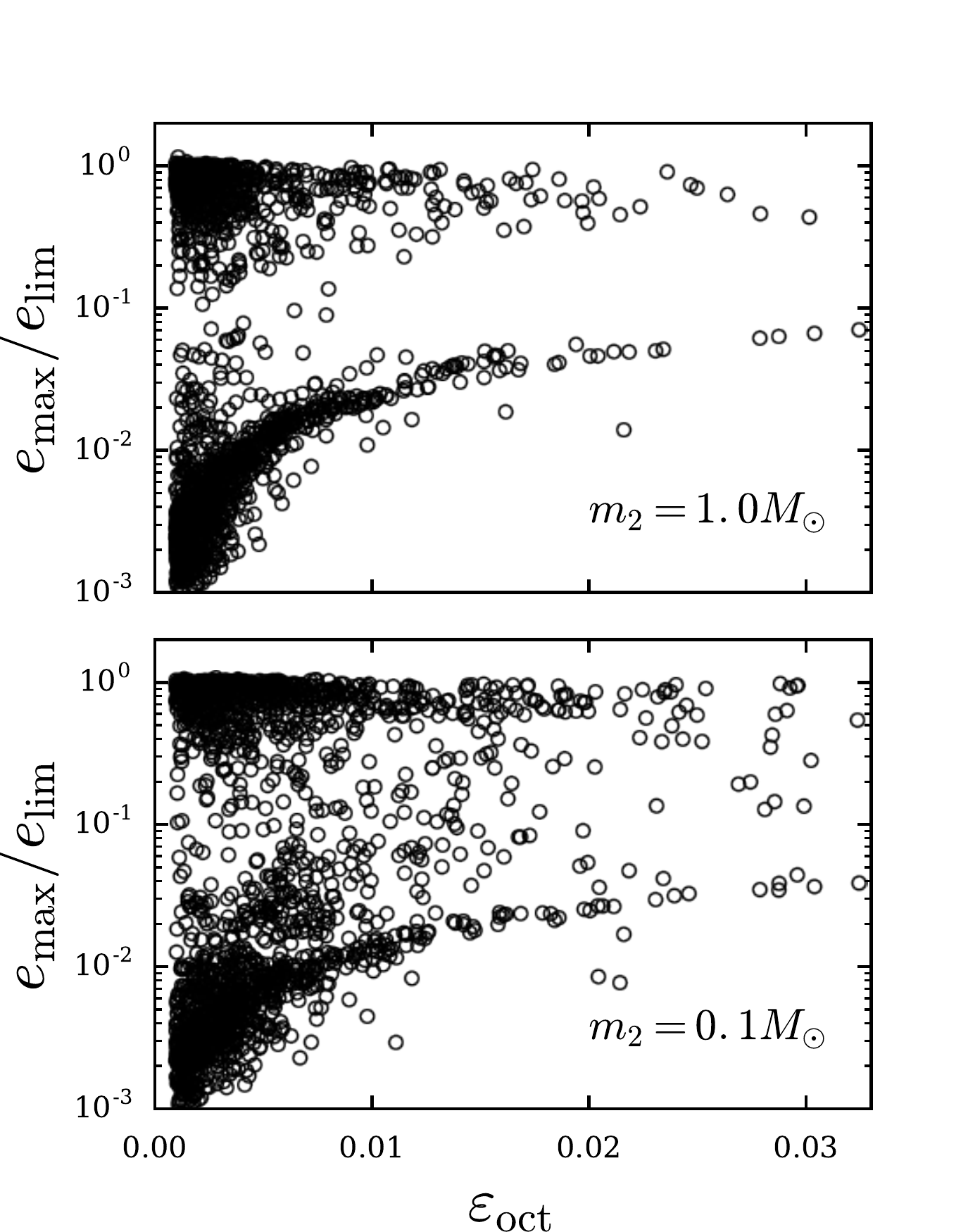}
\caption{Maximum eccentricity $\emax$ achieved over the integration timespan, compared to the analytically determined (quadrupole) limiting eccentricity $\elim$.  For a given value of $\varepsilon_\oct$, a wide range of $\emax/\elim$ is achieved depending on the chosen orbital configuration, but $\emax/\elim \leq 1$ is always satisfied.}
\label{fig:elim}
\end{figure}

To check whether this result is robust across a wide variety of systems, we repeat the previous Monte Carlo experiment conducted in Section \ref{subsec:equal} with $m_0 = 1 \msun$ and $m_1 = 0.5 \msun$.  All other parameters are sampled identically, with the additional selection criterion that $\varepsilon_\oct > 0.001$.  We integrate each system for $\sim 30 \tk /\varepsilon_\oct$, i.e. $\sim 30$ octupole LK timescales. In Fig.~\ref{fig:elim} we plot $\emax/\elim$ versus $\varepsilon_\oct$, where $\emax$ is the maximum eccentricity achieved over the entire numerical integration timespan, while $\elim$ is calculated from Eq.~(\ref{eq:jlim}).  Inspection of Fig.~\ref{fig:elim} reveals that $\emax \leq e_\Lim$. As a result, while knowledge of $\emax$ for an arbitrary inclination require a full numerical integration, the algebraic expression for the upper limit on $\emax$ (Eq.~[\ref{eq:jlim}]) remains valid for systems with non-zero octupole terms.

\begin{figure*}
\centering 
\includegraphics[width=0.8\textwidth]{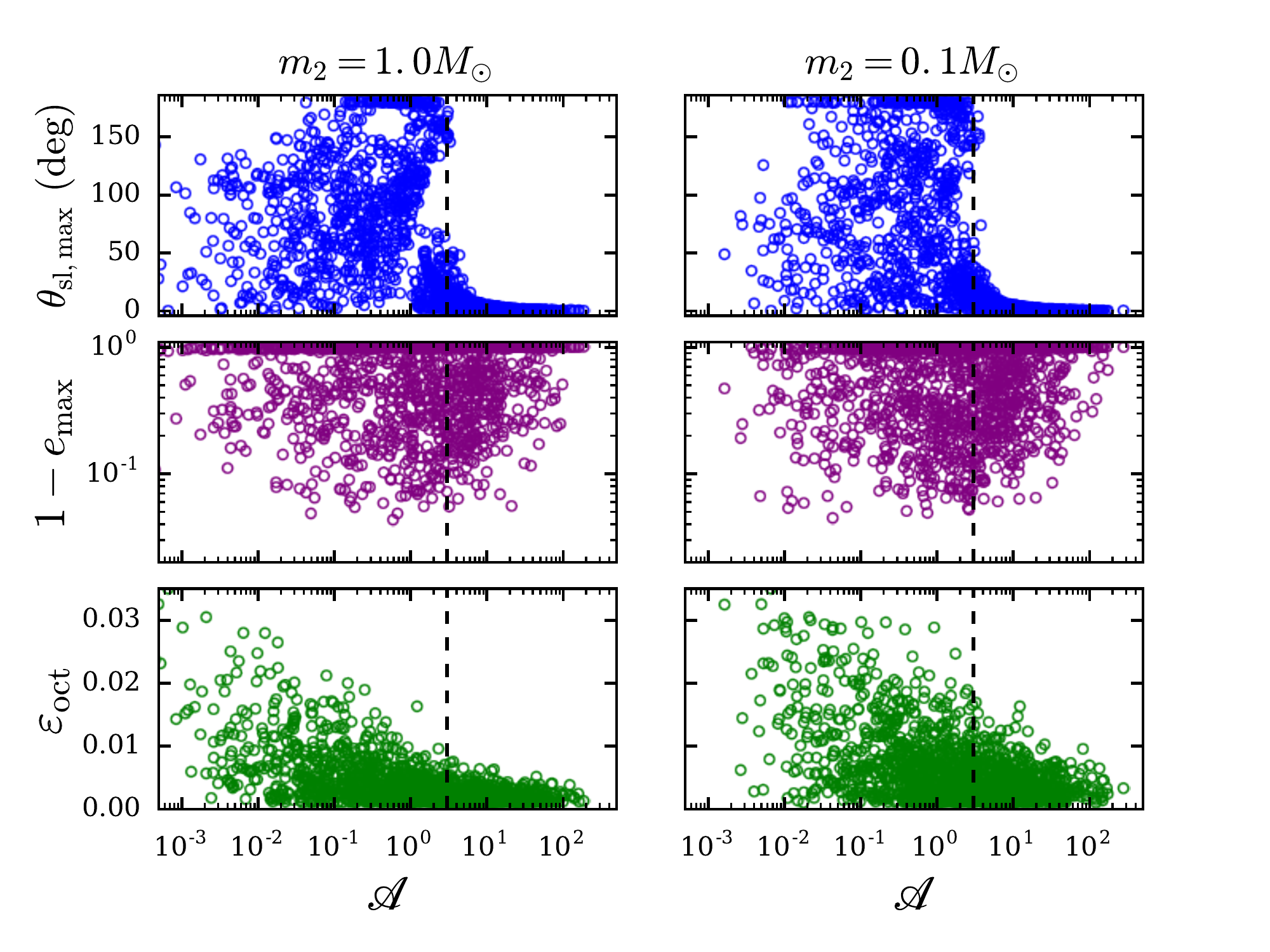}
\caption{Same experiment as depicted in Fig.~\ref{fig:Acrit_no_oct}, except that the inner binary has unequal mass ($m_0 = 1 \msun$, $m_1 = 0.5 \msun$), so that $\varepsilon_{\oct} \neq 0$ (shown in the bottom panel). Same set of simulations as in Fig.~\ref{fig:elim}. As found previously (for $\varepsilon_{\oct} = 0$), systems with $\A \gtrsim 3$ all maintain low spin-orbit misalignment for the entire integration.}
\label{fig:Acrit_with_oct}
\end{figure*}

Second, we examine whether the adiabaticity parameter $\A$ remains a useful quantity in predicting the ability to generate spin-orbit misalignment. The results are shown in Fig.~\ref{fig:Acrit_with_oct} (compare with Fig.~\ref{fig:Acrit_no_oct}).  We find again that systems with $\A > A_{\rm crit} \simeq 3$ all maintain low spin-orbit misalignment, while systems with $\A \lesssim 3$ do not.  A possible reason is that systems with the largest $\varepsilon_{\oct}$ tend to have $\A \lesssim 3$ (due to the strong dependence of $\A$ on $\aouteff$), and therefore lie in the non-adiabatic (low $\A$) regime.  As a result, octupole-level dynamics do not affect the existence or numerical value of $\mathcal{A}_{\rm crit}$, because the octupole contribution for systems near $\A_{\rm crit}$ is negligible.

To summarize Sections \ref{subsec:equal} and \ref{subsec:unequal}: By conducting a series of numerical integrations, with orbital parameters and stellar spin periods sampled over wide ranges, and considering both a solar-mass and brown dwarf tertiary companion, we have identified a condition necessary for generating substantial spin-orbit misalignment ($\theta_{\SL} \gtrsim 30^\circ$) of the inner binary.  The orbital geometries of the inner and outer binaries ($\ain,\aouteff$), and the stellar spin period ($P_\star$) must satisfy $\A \lesssim 3$, where $\A$ is given in Eq.~(\ref{eq:Aparameter}).  This result also holds for $\varepsilon_\oct \neq 0$.  We have also shown that the maximum achieved eccentricity of the inner binary in systems with octupole never exceeds the quadrupole limiting value, as determined by the algebraic expression in Eq.~(\ref{eq:jlim}).  Therefore, the analytical results in Section 2.3, constraining the maximum perturber distance capable of raising the eccentricity from $e \sim 0$ to $e_{\rm obs}$ (through LK oscillations), remain valid for $\varepsilon_\oct \neq 0$.

Taken together, these findings shed insight into the dynamical behavior of hierarchical stellar triples, without undertaking full numerical integrations. 

\section{Application: DI Herculis}
As an application of the results presented in previous sections, we consider the eclipsing binary DI Herculis.  DI Herculis consists of two B stars, with masses $m_0 = 5.15 \msun$ and $m_1 = 4.52 \msun$, orbital period $P \simeq 10.55$ days, and eccentricity $e \simeq 0.49$ \citep{popper1982}.  DI Herculis has been an object of interest, due to an observed pericenter precession rate too low compared with predicted rate due to general relativity \citep{martynov1980}. Both the primary and secondary components of this system were recently confirmed to have significant projected spin-orbit misalignments \citep{albrecht2009}, leading to an additional source of precession (with the opposite direction compared to GR), thereby resolving the anomalously low observed rate.  The projected obliquity of the primary is $\lambda_{\rm pri} \simeq 72^{\circ}$ and that of the secondary is $\lambda_{\rm sec} \simeq -84^{\circ}$.  Here, we consider the possibility that the large obliquities and eccentricity arose from secular perturbations from an undetected tertiary companion, and provide constraints that the hypothetical companion must satisfy.

\begin{figure}
\centering 
\includegraphics[scale=0.65]{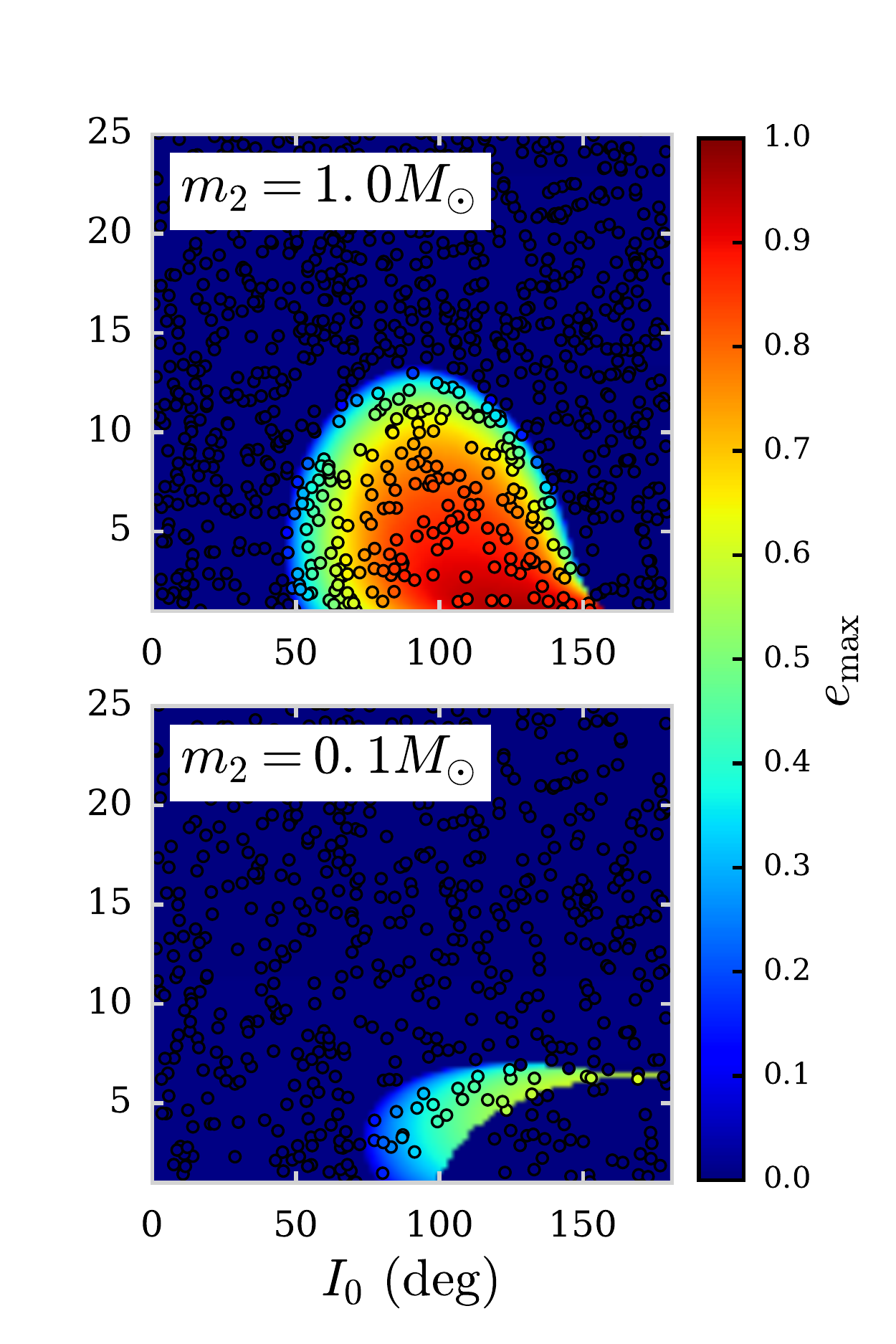}
\caption{Similar to Fig.~\ref{fig:aout_vs_I0}, but applied to the DI Herculis system, which has an inner binary with $m_0 \simeq 5.15 \msun$ $m_1 \simeq 4.52 \msun$, eccentricity $e_{\rm obs} \sim 0.5$, $P_{\rm orb} \simeq 10.55$ days, and estimated spin periods of $P_\star \sim 1$ day.  In order for the eccentricity of DI Herculis to have been increased from $\sim 0$ to $0.5$ by LK cycles from a tertiary companion, the LK maximum eccentricity must satisfy $\emax \geq e_{\rm obs} \simeq 0.5$. We show results for a stellar mass and brown dwarf perturber, as labeled, and have set $\eout = 0$ in this example.  The colored circles depict the results of numerical integrations of the full equations of motion, as discussed in Section \ref{sec:numerical}. The colormap depicts the analytic estimate of $\emax$ as discussed in Sections \ref{sec:window} and \ref{sec:emaxlim}.  In order to produce the observed eccentricity, a brown-dwarf perturber must be in a retrograde orbit.}
\label{fig:DIHercemax}
\end{figure}

If a tertiary companion is responsible for raising the eccentricity from $\sim 0$ to the observed value $e_{\rm obs} \simeq 0.5$, then the LK maximum eccentricity must satisfy $\emax \gtrsim 0.5$.  Considering ranges of inclinations and semi-major axes for hypothetical perturbers, the colormap in Fig.~\ref{fig:DIHercemax} shows the analytically-determined maximum eccentricity, calculated using the procedure described in Section 2.  To ensure that the analytic treatment properly captures the dynamics of DI Herculis, we have also undertaken full numerical integrations, depicted as solid circles.  In the analytic determination of $\emax$ (Section 2), we have considered the SRF contributions from GR, along with tidal and rotation distortion of both $m_0$ and $m_1$. In contrast to solar-type stars, effects of rotational distortion are important in both members of DI Herculis, because the large radii and rapid rotation rates lead to large rotation-induced quadrupole moments.  Recall that rotational distortion may only be incorporated in the analytic treatment of the LK maximum eccentricity in an approximate manner, and in Section 2 alignment of the rotation and orbital axes was assumed. A precise determination of $\emax$ thus requires full numerical integrations over a large number of LK cycles.  Despite the approximation of aligned spin and orbital axes, the analytic treatment is in near perfect agreement with results from numerical integrations.

Inspecting Fig.~\ref{fig:DIHercemax}, a solar-mass perturber must be located within $\sim 12$ AU, with a wide range of possible inclinations.  In contrast, the required properties of a brown dwarf perturber are much more restrictive.  A brown dwarf perturber must be located within $\sim 5$ AU in a retrograde orbit.  Different choices for the outer binary's eccentricity will modify these constraints.  However, given that $m_0 \simeq m_1$, the DI Herculis system is unlikely to be significantly affected by octupole contributions, so the perturber's eccentricity can be absorbed into the definition of the ``effective'' semi-major axis $\aouteff = \aout \sqrt{1 - \eout^2}$ (unless the angular momentum ratio satisfies $\eta \gtrsim 1$).

\begin{figure}
\centering 
\includegraphics[scale=0.55]{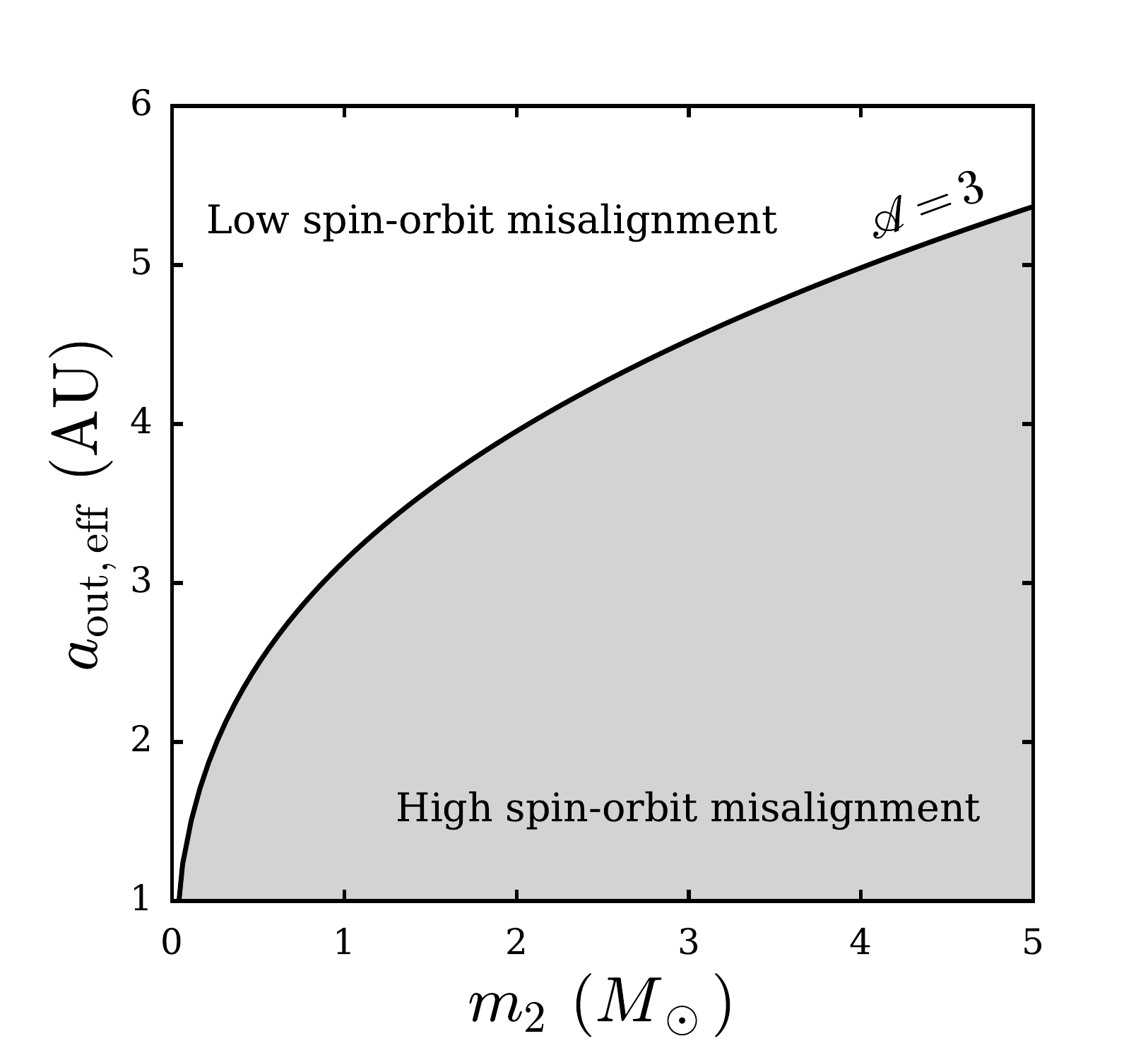}
\caption{Required effective separation $\aouteff = \aout \sqrt {1 - \eout^2}$ versus mass of a tertiary companion $m_2$ in the DI Herculis system, to generate the large inferred spin-orbit misalignment of the primary member.  In order to produce the misalignment, the inner and outer binaries must satisfy $\A \lesssim 3$, as indicated by the shaded region.  As a result, the outer binary must have $\aouteff \lesssim 1-5$ AU, depending on the tertiary mass.  Note that the minimum semi-major axis necessary for stability is \citep[$\sim 1$ AU,][]{mardling2001}.}
\label{fig:DIHerc}
\end{figure}

If a tertiary companion is responsible for raising the spin-orbit angle of either member of DI Herculis from $\sim 0$ to the observed values, the adiabiaticity parameter must satisfy $\A \lesssim 3$ (see Sections 3 and 4, and Eq.~[\ref{eq:Aparameter}]).  The rapid rotation rates of both stars ($v \sin i > 100$ km s$^{-1}$), combined with the large stellar radii, implies that a perturber must be extremely close and/or massive to achieve $\A \lesssim 3$.  Figure \ref{fig:DIHerc} depicts the combinations of $m_2$ and $\aouteff$ that lead to $\A < 3$ for the primary member (shaded region).  Note that we have assumed a primary stellar spin period $P_\star = 1.25$ days -- this rapid rotation rate is consistent with the observed $v \sin i$, and the estimated value by \cite{philippov2013} using gravity darkening.  Inspecting Fig.~\ref{fig:DIHerc}, we see that a perturber with $m_2 \sim 1 \msun$ must have an effective separation $\aouteff \lesssim 3$ AU, and if $m_2 \sim 0.1 \msun$, $\aouteff \lesssim 1$ AU.  Note that such triple systems are only marginally stable -- the \cite{mardling2001} stability criterion (see Eq.~\ref{eq:stability}) yields a minimum separation of $\aout \sim 1$ AU.  

The requirement that a solar-mass perturber be located within $\sim 3$ AU in order to generate the observed spin-orbit misalignment may be problematic, given that no additional bodies have been observed.  A low-mass (e.g. brown dwarf) perturber is much more compelling than a solar-mass perturber, because it is more likely to have hitherto escaped detection.  However, the requirement that it be located within $\sim 1$ AU) would place it uncomfortably close to the stability limit.  

To summarize: we have considered the possibility that the observed eccentricity and/or spin-orbit misalignment in the DI Herculis system result from secular perturbations from a tertiary companion.  The perturber must be located within $\sim 5 - 15$ AU to generate the observed eccentricity.  The constraints on based on the obliquity are much more stringent, and the perturber must be located within $\sim 1 - 3$ AU (depending on perturber mass), very close to the stability limit.

\section{Conclusion}
\subsection{Summary of Key Results}
This paper has examined the secular dynamics of hierarchical stellar triples, with the goal of identifying the requirements for the tertiary body to induce spin-orbit misalignment and/or eccentricity in the inner binary through Lidov-Kozai cycles in concert with stellar spin-orbit coupling.  We have considered the orbital evolution of both the inner and outer binaries, combined with the dynamics of mutual spin-orbit nodal precession, as well as pericenter precession from various short range-forces (SRFs), such as general relativity and stellar tides.  The results of this paper allow constraints to be placed on hidden tertiary companions in stellar binaries that exhibit spin-orbit misalignment or eccentricity.  The key results of this paper are: 

$\bullet$  We derive new analytic expressions for the maximum eccentricity achieved by the inner binary (Eq.~[\ref{eq:energy}]) and the ``LK window'' for eccentricity excitation (see Eqs.~[\ref{eq:cos_range}], [\ref{eq:Imin}], and Fig.~\ref{fig:window}) due to the secular quadrupolar forcing of an external companion.  The quadrupole approximation is exact when the inner binary has equal masses, or when the outer binary orbit is circular.  Note that these calculations assume an initial inner binary eccentricity $e_0 \simeq 0$.  Our expressions for $\emax$ and the LK window are valid for general masses of the triples and include the effects of SRFs. Our analysis generalizes previous work, which was restricted to small mass ratios and/or neglect SRFs.  These results shed light on the dynamical behaviors of hierarchical triples with a wide range of parameters, without undertaking full numerical integrations.

$\bullet$  For an observed binary system with eccentricity $e_{\rm obs}$, constraints on unseen tertiary companions can be made, by requiring $\emax \ge e_{\rm obs}$, assuming the inner binary has an initial eccentricity $e_0 \simeq 0$.  Although the assumption $e_0\simeq 0$ may not always hold  (since the inner binary may form with a range of eccentricities), this constraint ($e_{\rm obs} \le \emax$) nonetheless provides useful information on the required masses and separation of an undetected tertiary companion (see Section \ref{sec:hidden}).  See Appendix \ref{sec:appendix} for a brief consideration of $e_0 \neq 0$. 

$\bullet$  In cases where the octupole contribution is important (when $m_0\neq m_1$ and $\eout \neq 0$), we carry out numerical experiments to determine $\emax$ (see Figs.~\ref{fig:elim_example} and \ref{fig:elim}).  As first noted by \cite{liu2015}, and confirmed in this paper under general conditions (e.g., arbitrary mass ratios for the hierarchical triples), the maximum eccentricity (with octupole) never exceeds the analytic quadrupole limiting eccentricity $\elim$ (see Section \ref{sec:emaxlim}, Eq.~[\ref{eq:jlim}]).  Without octupole this limiting eccentricity is only achieved ($\emax = \elim$) at a specific value of the initial inclination $I_{0,\Lim} \geq 90^\circ$ (see Eq.~[\ref{eq:Ilim}]), but including octupole allows $\emax = \elim$ to be realized for a wider range of inclinations.  Since $\emax \leq e_\Lim$, constraints can be placed on the required perturber properties ($m_2,\aout,\eout$) needed to generate eccentricity even in systems with octupole contributions, without undertaking numerical integrations.  

$\bullet$ From numerical integration of the full secular equations of motion (including the dynamics of the orbits and stellar spins) for a variety of triples and stellar parameters, we have identified a robust, necessary condition for generating spin-orbit misalignment in the inner binary due to perturbations from a tertiary companion: Large ($\gtrsim 30^\circ$) misalignment can be generated only if the stellar masses, spin period, and the orbital parameters of the triples are such that the ``adiabaticity parameter'' $\A$, defined by Eq.~(\ref{eq:Aparameter}), satisfies $\A \lesssim 3$ (see Figs.~\ref{fig:Acrit_no_oct} and \ref{fig:Acrit_with_oct}).  

Physically, $\A$ is the ratio of the precession rate of the stellar spin (driven by the secondary) and the orbital precession rate of the inner binary (driven by the tertiary), evaluated at inner binary zero eccentricity.  This finding ($\A \lesssim 3$ for producing misalignment) holds across wide ranges of orbital architectures and stellar spin periods.  Although theoretical work on spin-orbit dynamics in binaries undergoing LK oscillations shows that the dynamics of the spin axis depends on more than a single parameter \citep{storch2016}, we find empirically that $\A \lesssim 3$ is highly effective in predicting whether large spin-orbit misalignment will occur, and has the advantage that it is easily evaluated for observed binaries.  For a specified inner binary separation, $\A \lesssim 3$ translates into the requirement that the outer binary must have a small effective separation $a_{\Out,\eff} = \aout \sqrt{1 - \eout^2}$, and/or the stellar rotation period must be short (see Fig.~\ref{fig:parameters_no_oct}).   Although the main focus of this paper has been on inclined tertiary companions, we note that nearly coplanar tertiaries can also increase spin-orbit misalignment and eccentricity, provided that the outer orbit is sufficiently eccentric, and the adiabaticity parameter satisfies $\A \sim 3$.  

$\bullet$ In Section 5 we apply our general results to the eclipsing binary system DI Herculis, and identify the properties that an undetected tertiary companion must satisfy, in order to be responsible for the observed eccentricity and spin-orbit misalignments.

\subsection{Discussion}
As noted in Section 1, this paper has neglected the effects of tidal dissipation in the inner binary.  Therefore, when applying our results (analytic expressions and various constraints) to observed binaries, it is important to make sure that the system under consideration has a sufficiently large pericenter distance so that its eccentricity and spin-orbit misalignment angle have not been affected by tidal dissipation within the lifetime of the system.  

Another physical effect ignored in this paper is stellar spin-down by magnetic braking.  Our pevious works \citep{storch2014, kra2016} have shown that stellar spin-down can significantly influence the final spin-orbit mislaignments of hot Jupiter systems formed through Lidov-Kozai migration.  Although the integration timespans considered in this paper have been sufficiently short so that $P_\star = $ constant is a valid approximation, the decrease in the stellar spin rate over Gyr timescales could be significant (depending on stellar type), and can reduce $\A$ by $\sim 10$ for solar-mass stars.  As stellar spindown takes place, the adiabaticity parameter may cross $\A \sim 3$, so that substantial misalignment is generated only late in the binary's evolution.  As a result, stellar triples where tidal decay does not occur in the inner binary may exhibit an increase in spin-orbit misalignment with stellar age.  

As noted above, the analytic results presented in this paper are valid for hierarchical triples with arbitrary masses.  Thus, they also have applications in exoplanetary systems consisting of two well-separated planets.  While numerous planets within $\sim 1$ AU of their host stars have been discovered from both transit and radial velocity searches, detection of more distant planets has proceeded more slowly.  Many observed planets within 1 AU have substantial eccentricities, and a possible explanation is secular interactions with additional undetected distant planets.  In systems containing an eccentric planet, the method developed in this paper can be used to place constraints on additional external planetary companions.  We plan to study these issues in a future paper.

\section*{Acknowledgments}
We thank the referee, Alexandre Correia, for useful comments. This work has been supported in part by NASA grants NNX14AG94G and NNX14AP31G, and a Simons Fellowship from the Simons Foundation.  K.R.A. is
supported by the NSF Graduate Research Fellowship Program under Grant
No. DGE-1144153.  N.I.S. is supported by a Sherman Fairchild Fellowship at Caltech. 

\appendix
\section{LK Maximum Eccentricity for Non-zero Initial Eccentricity}
\label{sec:appendix}
In this Appendix, we demonstrate how the analytic results of Section 2 may be modified when the initial eccentricity $e_0 \neq 0$.  In the following results, we restrict the initial eccentricity to moderate values, $e_0 \lesssim 0.3$.  This is justified because our goal is to identify the required properties of tertiary companions in raising the eccentricity of binaries starting from low or moderate initial values.

For general values of the initial eccentricity $e_0$, $e$ oscillates between a minimum value $\emin$ and a maximum value $\emax$, with $\emin \leq e_0 \leq \emax$.  Both $\emax$ and $\emin$ depend on the initial pericenter angle $\omega_0 \equiv \omega(e_0)$.  If $\omega_0 = 0, \pi$ or $\omega_0 = \pi/2, 3 \pi/2$, then either $e_0 = \emin$ or $e_0 = \emax$.  For other values of $\omega_0$, we have $\emin \leq e_0 \leq \emax$.

When $e_0 \neq 0$, the minimum and maximum eccentricities may occur either at $\omega = 0, \pi$ or $\omega = \pi/2, 3 \pi/2$, and $\omega$ may either circulate or librate.  To determine $\emax$ from a given set of initial conditions, we calculate $\omega(e)$ using energy conservation, given by:   
\be
\Phi_{\rm Quad} (e, \omega) + \Phi_{\rm SRF} (e) = \Phi_{\rm Quad} (e_0, \omega_0) + \Phi_{\rm SRF} (e_0).
\ee
See Section \ref{sec:setup} for definitions of $\Phi_{\rm Quad}$ and $\Phi_{\rm SRF}$. Requiring $0 \leq \cos^2 \omega \leq 1$ allows the maximum and minimum eccentricities to be determined, and are given by ${\rm max}[e(\omega)]$ and ${\rm min}[e(\omega)]$.

\begin{figure*}
\centering 
\includegraphics[width=\textwidth]{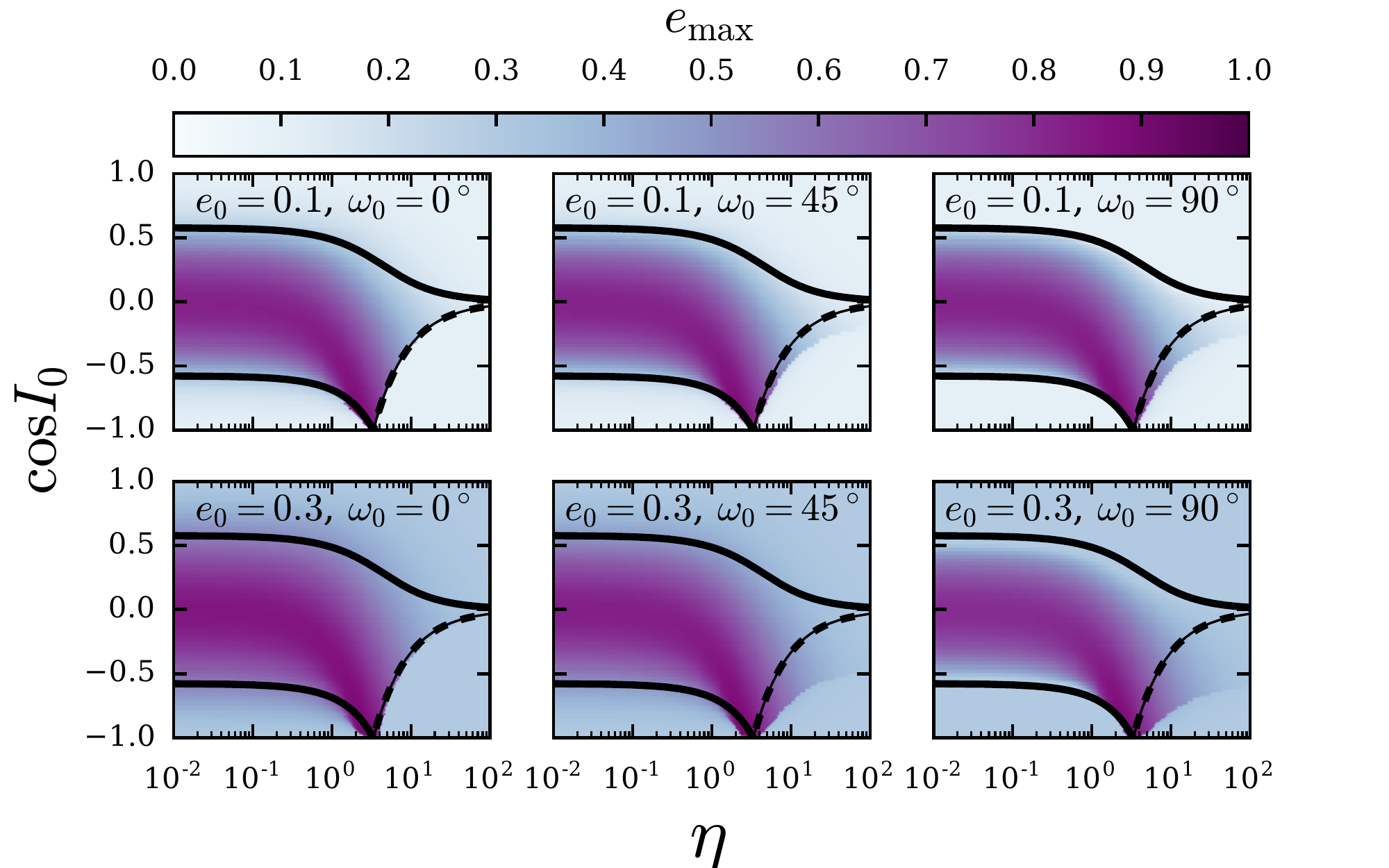}
\caption{$\emax$, in terms of $\eta$ and $\cos I_0$, for various combinations of $e_0$ and $\omega_0$.  We have fixed $\varepsilon_{\gr} = 1$, and have set $\varepsilon_{\tide}$, $\varepsilon_{\rm Rot} = 0$. Compare with Fig.~\ref{fig:window}.  For reference, the black curves show the analytic expressions for the range of $\cos I_0$ allowing eccentricity increases from $e_0 \simeq 0$ (the ``LK window''), derived in Section \ref{sec:window} (Eqs.~[\ref{eq:Icrit}] and [\ref{eq:Imin}]).  Non-zero $e_0$ does not substantially modify the LK window unless $\eta \gtrsim 1$.}
\label{fig:window_general_e0}
\end{figure*}

For specified $(e_0,\omega_0)$, along with the orbital geometry and physical properties of $m_0$, $m_1$, and $m_2$ (which enter through $\eta$, $\varepsilon_{\gr}$, $\varepsilon_{\tide}$ and $\varepsilon_{\rm Rot}$; see Eqs.~[\ref{eq:eta}] and [\ref{eq:epsilon_SRFs}]), the value of $\emax$ depends on the initial inclination $I_0$.  In the case of $e_0 \simeq 0$, the ``LK window,'' (i.e. the range of inclinations that allow eccentricity oscillations) may be explicitly calculated (see Section \ref{sec:window}), and takes the simple form of Eqs.~(\ref{eq:Icrit}) and (\ref{eq:Imin}).  When $e_0 \neq 0$, the LK window is modified, and becomes somewhat fuzzier. In Fig.~\ref{fig:window_general_e0} we demonstrate how non-zero $e_0$ affects the LK window, by calculating $\emax$ as function of $\eta$ and $\cos I_0$, for a fiducial value of $\varepsilon_{\gr}$ and several different combinations of ($e_0,\omega_0$).  Compare with Fig.~\ref{fig:window}.  For reference, the explicit expressions for the LK window when $e_0 \simeq 0$ (Eqs.~[\ref{eq:Icrit}] and [\ref{eq:Imin}]) are also shown. For $\eta \lesssim 1$, Eqs.~(\ref{eq:Icrit}) remain an excellent prediction of whether eccentricity excitation may occur, regardless of the values of $e_0$ and $\omega_0$.  When $\eta \gtrsim 1$ and $\omega_0 \neq 0$, the range of inclinations allowing eccentricity increases is modified compared to the $e_0 \simeq 0$ case.

\begin{figure*}
\centering 
\includegraphics[width=\textwidth]{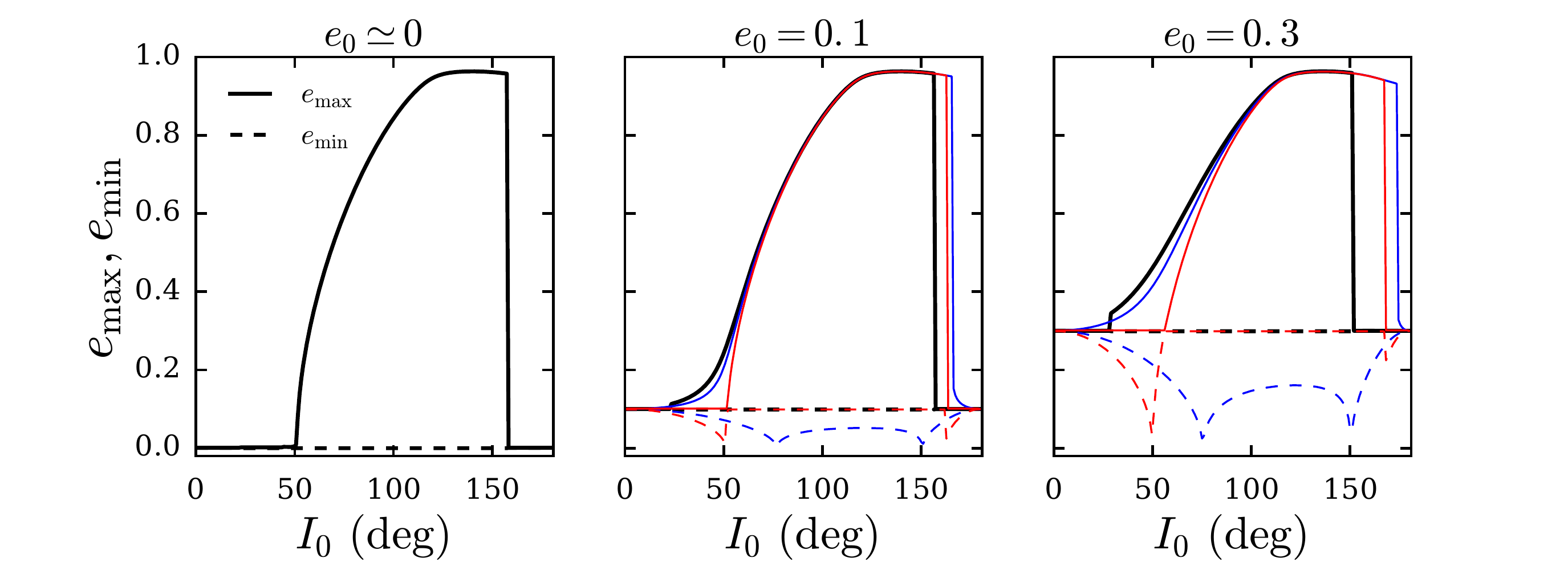}
\caption{Maximum and minimum eccentricities as a function of initial inclination, for various initial eccentricities $e_0$ and phase angles $\omega_0$.  We show $\omega_0 = 0^\circ$ (black curves), $\omega_0 = 45^\circ$ (blue curves), and $\omega_0 = 90^\circ$ (red curves).  The solid curves depict $\emax$ and the dashed curves depict $\emin$. Compared to the $e_0 \simeq 0$ case, non-zero $e_0$ can lead to eccentricity oscillations for a wider range of $I_0$, depending on the value of $\omega_0$.  The lower inclination boundary for eccentricity growth $(\cos I_0)_+$ approaches zero, but the upper boundary corresponding to $(\cos I_0)_-$ remains. The orbital and physical parameters are the same as in Fig.~\ref{fig:emax}.  $I_{\Lim}$ and $e_{\Lim}$ (see Section \ref{sec:emaxlim}) are nearly independent of $e_0$ and $\omega_0$.}
\label{fig:emax_general_e0}
\end{figure*}

Figure \ref{fig:emax_general_e0} depicts $\emax$ and $\emin$ versus $I_0$ for several different values of $e_0$ and $\omega_0$, assuming the same orbital and physical parameters as in Fig.~\ref{fig:emax}.  As discussed in Section \ref{sec:emaxlim}, there is a value of $I_0$ that yields a maximum value of $\emax$ (the ``limiting eccentricity''), denoted as $I_{0,\Lim}$ and $\elim$ respectively. Regardless of $e_0$ and $\omega_0$, $\elim$ and $I_{0,\Lim}$ have nearly the same values. 

\end{document}